\documentclass[aap,preprint]{imsart}
\usepackage{amsmath}
\usepackage{pdflscape}
\usepackage{graphicx,psfrag,epsf}
\usepackage[colorlinks,breaklinks,citecolor=blue,urlcolor=blue]{hyperref}
\usepackage{natbib}
\setattribute{journal}{name}{}  %% With imsart class, can suppress text Submitted to...

\arxiv{1810.11557}

\defcitealias{MazalovPeshkov2004}{MP (2004)}
\def\spacingset#1{\renewcommand{\baselinestretch}%
{#1}\small\normalsize} \spacingset{1}

\defcitealias{GilbertMosteller1966}{GM (1966)}
\def\spacingset#1{\renewcommand{\baselinestretch}%
{#1}\small\normalsize} \spacingset{1}

\defcitealias{PresmanSonin1972}{PS (1972)}
\def\spacingset#1{\renewcommand{\baselinestretch}%
{#1}\small\normalsize} \spacingset{1}

\begin{document}

\begin{frontmatter}

\title{The Duration of Optimal Stopping Problems}
%% When Should People Get Married?
%% A  Note  on  the  Number  of Interviews  When  Talent  is Uniformly  Distributed
%% \protect\thanksref{T1}
\runtitle{Duration of Optimal Stopping Problems}
%%\thankstext{T1}{Footnote to the title with the `thankstext' command.}

\begin{aug}
  \author{Simon Demers\ead[label=e1]{simon.demers@vpd.ca}}
%%  \thankstext{t2}{}
  \runauthor{S. Demers}
  \affiliation{}
  \address{Vancouver Police Department \\ 3585 Graveley Street \\ Vancouver, British Columbia, Canada \\V5K 5J5\\ \printead{e1}}
\end{aug}

\begin{abstract}
Optimal stopping problems give rise to random distributions describing how many applicants the decision-maker will sample or interview before choosing one, a quantity sometimes referred to as the search time or process duration. This research note surveys several variants of optimal stopping problems, extends earlier results in various directions, and shows how many interviews are expected to be conducted in various settings. The focus is on problems that require a decision-maker to choose a candidate from a pool of sequential applicants with no recall, in the vein of previously studied Cayley-Moser, Secretary and Sultan's Dowry problems.
\end{abstract}



\end{frontmatter}

\section{Introduction}

Although optimal stopping problems have been studied and refined extensively over time \citep{GilbertMosteller1966,Freeman1983,Ferguson1989}, relatively little is currently known about the statistical properties of the random distributions they give rise to. For example, it is not obvious how the expected number of observations (``applicants'') to be considered (``interviewed'') before one is ultimately chosen (``hired'') will vary when the decision-maker (``employer'') can observe the actual payoff values (``talent'') or when the horizon (``applicant pool'') grows. This is unfortunate because it might be useful in some practical settings to anticipate, for example, how many interviews the decision-maker should plausibly schedule or prepare for. This is what this research note attempts to elucidate.

The focus is on the problem of choosing a candidate from a pool of applicants with no recall and uniformly distributed talent. We first provide some background information about the problem and summarize useful earlier results (Section~\ref{Sec2}). We then derive algebraically the mean and median number of interviewed applicants expected within the no-information setting (Section~\ref{Sec3}), where the decision-maker can only assess applicants using a rank-based indicator \citep{Bearden2006} or only cares about choosing the best candidate anyway, as in the classical Secretary Problem or the Sultan's Dowry Problem \citep{Mosteller1987}. By leveraging earlier results from \cite{Yeo1997}, we then generalize the analysis to no-information rank problems whose solution is characterized by a series of multiple thresholds (Section~\ref{Sec4}). This includes the \cite{GuseinZade1966} setting and the minimum rank problem of \cite{Lindley1961}. Finally, we turn to the full-information settings of \cite{Moser1956} and \citet[\S 3]{GilbertMosteller1966}, where payoff values are observable (Section~\ref{Sec5}).

\section{Background and Known Results}\label{Sec2}

Consider the problem of a decision-maker (``employer'') who is looking for the best possible candidate (``hire'') out of a random sequence of $N$ applicants sampled from a uniform distribution but cannot recall previously considered applicants who have been passed on.

\subsection{Full Information, Cardinal Payoffs}

Assume for now that the decision-maker's payoff value is determined by the selected candidate's attractiveness, quality, or intrinsic value (interchangeably) and that the decision-maker can actually observe each sequential applicant's attractiveness (``payoff value''). This is the full-information setting of \cite{Moser1956}.

Let $X_i$ be the payoff associated with the $i\textsuperscript{th}$ applicant. Assume that the observations $X_1,X_2,\dots,X_N$ are independent and identically distributed, drawn from a known uniform distribution scaled on the interval $[0,1]$.

As a reminder, when there are $m=N-i$ applicants left to be observed (``interviewed''), the optimal stopping rule consists in stopping and choosing the $i\textsuperscript{th}$ applicant whenever $X_i > A_m$ \citep{Moser1956}, where $A_m$ is defined recursively (inductively) with $A_0=0$, $A_1=0.5$ and
\begin{equation}\label{eqMoserAm}
    A_{m+1} = \frac{{A_m}^2+1}{2}.
\end{equation}

When the pool of applicants left to be observed is large enough, the cutoff points can be approximated as: $A_m \simeq 1 - 2/[m+\ln(m)+1.76799]$ \citep[Eq.~5a-7]{GilbertMosteller1966}. Of interest, \cite{MazalovPeshkov2004} previously proved that the expected number of interviews in that setting converges asymptotically to $N/3$.

\subsection{No Information, Cardinal Payoffs}

Fifty years after \cite{Moser1956}, \cite{Bearden2006} considered a similar problem but assumed instead that the decision-maker could only observe an indicator revealing whether each applicant is the relatively best observed so far, in the true tradition of the Secretary Problem \citep{Ferguson1989}. This is the no-information setting with cardinal payoffs.

As shown by \cite{Bearden2006}, the optimal strategy for the decision-maker who can only observe whether each successive applicant is the relatively best so far but cannot observe the actual payoff values is to reject the first $\sqrt{N}-1$ applicants (rounded to the nearest integer) and select the next candidate identified as the relatively best so far, or the $N\textsuperscript{th}$ applicant if none turns out to be relatively better than the $\sqrt{N}-1$ applicants observed initially.

The reasoning behind the threshold rule obtained by \cite{Bearden2006} is as follows. The decision-maker only observes an indicator $I_i$, where $I_i=1$ if and only if the $i\textsuperscript{th}$ applicant is the relatively most attractive (best) so far and $I_i=0$ otherwise. Let $T$ be the number of applicants interviewed before one is hired. Given a pool of $N$ applicants and an arbitrary threshold of $c$ applicants with $1 \leq c \leq N$, the probability that the $x\textsuperscript{th}$ applicant will be chosen is:
\begin{equation}\label{eqProbDistT}
	\Pr(T=x) = 
	\begin{cases}
		\displaystyle \left[ \prod_{s=c}^{x-1} \left( \frac{s-1}{s} \right) \right] \frac{1}{x} = \frac{c-1}{x(x-1)} & \text{for}\ c<x<N \\
		\displaystyle \prod_{s=c}^{N-1} \left( \frac{s-1}{s} \right) = \frac{c-1}{N-1} & \text{for}\ x=N \\
		0 & \text{otherwise.}\
	\end{cases}
\end{equation}

Given that the $x\textsuperscript{th}$ applicant is relatively best, its expected payoff value is $\operatorname{\mathbf{E}}(X_x \, | \, I_x=1) = x/(x+1)$. If the decision-maker is instead compelled to select the last ($N\textsuperscript{th}$) applicant by default, the expected value is simply the unconditional mean: $\operatorname{\mathbf{E}}(X_N)=0.5$. Combining these arguments, \cite{Bearden2006} showed that the expected value for the decision-maker is:
\begin{align}\label{eqBearden2006EV}
	EV(c) & = \sum_{x=c}^{N-1} \left[ \prod_{s=c}^{x-1} \left( \frac{s-1}{s} \right) \right] \left( \frac{1}{x+1} \right) + 0.5 \prod_{s=c}^{N-1} \left( \frac{s-1}{s} \right) \nonumber \\
	& = \sum_{x=c}^{N-1} \left( \frac{c-1}{x-1} \right) \left( \frac{1}{x+1} \right) + 0.5 \left( \frac{c-1}{N-1} \right) \nonumber \\
	& = \frac{2Nc-c^2+c-N}{2Nc}
\end{align}

It is a matter of algebra to show that this expected value is maximized by setting $c^*=\sqrt{N}$. That is, the optimal strategy for the decision-maker is to reject the first $\sqrt{N}-1$ applicants (rounded to the nearest integer) and select the next candidate identified as the relatively best so far. Under this optimal selection strategy, the decision-maker's expected value is $EV(c^*)=1-(2\sqrt{N}-1)/(2N)$.

\subsection{No Information, Best Choice}

When the decision-maker cannot observe the payoff values but only cares about choosing the best applicant anyway, maximizing the expected payoff becomes equivalent to maximizing the probability of selecting the best applicant. This case is sometimes referred to as the Sultan's Dowry Problem \citep{Mosteller1987} but it embodies all the essential elements of the classical Secretary Problem \citep{Ferguson1989}.

Given an arbitrary threshold rule with threshold $c$, the probability that the best applicant will be chosen is:
\begin{align}\label{eqSultanPr}
    \Pr(\text{Success})
%%    & = \sum_{k=c}^{N} \Pr(k\textsuperscript{th} \text{ applicant is selected and it is the best})\nonumber\\
%%    & = \sum_{k=c}^{N} \Pr(k\textsuperscript{th} \text{ applicant best}) \Pr(k\textsuperscript{th} \text{ applicant selected}  \, | \, \text{best}) \nonumber\\
    & = \frac{1}{N} \sum_{k=c}^{N} \frac{c-1}{k-1} = \frac{c-1}{N} \sum_{k=c}^{N} \frac{1}{k-1}
\end{align}

The optimal threshold value $c^*$ that maximizes Eq.~\eqref{eqSultanPr} is the first $c$ such that $\Pr(c+1) \leq \Pr(c)$, which reduces to $\sum_{c+1}^{N} 1/(k-1) \leq 1$. The optimal thresholds for applicant pools of size $N \leq 1000$ are given by the integer sequence A054404 in the \cite{OEISA054404}, while useful algebraic approximations are given by \cite{Weisstein2004}.

Asymptotically, it is well known that the optimal threshold tends towards $c^* \sim N/\mathrm{e}$ and, equivalently, the proportion of applicants who will be sampled but automatically passed over tends to $c^*/N \sim \mathrm{e}^{-1} \approx 0.368$. Incidentally, the probability of successfully choosing the single best applicant also tends towards $\mathrm{e}^{-1}$ as the pool size becomes very large \citep{Ferguson1989}.

The heuristic argument in the asymptotic case can be made simple by recognizing that the decision-maker in the classical Secretary Problem or Dowry Problem will be satisfied by nothing but the best and therefore will only consider candidates who are identified as relatively best so far (with relative rank 1 or $I_i=1$). The \emph{ex ante} probability that the $i\textsuperscript{th}$ applicant will be relatively best is $1/i$. Taking the sum over the first $n$ applicants gives the expected number of relatively best candidates: $H_n = \sum_{1}^{n} 1/i$, which is also the $n\textsuperscript{th}$ harmonic number. For asymptotically large $n$, the first-order approximation is $H_n \sim \ln{(n)}+\gamma$, where $\gamma = 0.5772\dots$ denotes the Euler-Mascheroni constant. With a total of $N$ applicants and an arbitrary threshold rule that consists of passing over the first $c$ applicants, the number of relatively best candidates expected to be found after the threshold has been passed therefore asymptotically tends to $\sum_c^N 1/i = H_N - H_c \sim \ln{(N)} - \ln{(c)} = \ln{(N/c)}$.

Obviously, if zero candidates are found after the threshold $c$ has been passed, it means that the absolute best applicant was within the first $c$ applicants and the decision-maker will lose. On the other hand, if more than one candidate is available to be selected after the threshold has been passed, then the decision-maker will select the first one and will lose because the next relatively best candidate in the sequence is necessarily better, by definition. The decision-maker will therefore only win if there is one and only one relatively best candidate waiting to be selected after the threshold is passed. The decision-maker therefore desires to set $\ln{(N/c)}=1$. This gives the desired asymptotic result: $c^*=N/\mathrm{e}$.

\subsection{Full Information, Best Choice}\label{secFIBC}

\citet[\S 3]{GilbertMosteller1966} studied the optimal stopping rule and expected success rate of the decision-maker who can observe the payoff values but only cares about choosing the best applicant. This case is the full-information, best-choice problem \citep{Gnedin1996}.

In that setting, when there are $m=N-i$ applicants left to be observed (``interviewed''), the optimal stopping rule consists in stopping and choosing the $i\textsuperscript{th}$ applicant whenever this applicant is the relatively best observed so far and it has value $X_i > A_m$, where $A_0=0$ and $A_{m}$ is the solution to $\sum_{j=1}^{m} ({A_m}^{-j}-1)/j = 1$ for $m=1,2,\dots$ \citep{GilbertMosteller1966,Gnedin1996}.

Asymptotically, when the pool of applicants left to be observed is large enough, the cutoff points are approximately $A_m \simeq 1 - c/m$, where $c = 0.804352\dots$ is the solution to:
\begin{equation}\label{eqc}
    \sum_{j=1}^{\infty} \frac{c^{j}}{j!\,j} = c + \frac{c^2}{2!2} + \frac{c^3}{3!3} + \dots = 1.
\end{equation}  %% Wolfram Alpha: 0.804352262845638

The probability of success decreases with $N$ but never falls below the lower bound:
\begin{equation}\label{eqPrSuccessGM1966}
    \Pr(\text{Success}) \simeq \frac{1}{\mathrm{e}^c} + \left( \mathrm{e}^c-c-1 \right) \int_1^\infty \frac{1}{x \cdot \mathrm{e}^{cx}} dx = 0.580164\dots
    %% \lim_{N \to \infty} \Pr(\text{Choosing Best}) =
\end{equation}

As in the no-information best-choice setting, this limiting success probability does not depend on the actual distribution of talent \citep{Samuels1993}, because the relative ranks and the best choice in particular remain insensitive to monotone transformations along the real line. In best-choice settings, any non-uniform distribution of talent can be re-scaled to achieve a uniform distribution that is monotonically equivalent.

\subsection{No Information, Best S or Better}\label{BestS}

One way to generalize the classical best-choice problem is to consider that the decision-maker is successful as long as the selected candidate has an overall rank of $S$ or better out of the $N$ available applicants. This is the Gusein-Zade Problem \citep{GuseinZade1966,FrankSamuels1980}. As usual, it is assumed in the no-information setting that the decision-maker can only observe the (relative) rank of each applicant among those who have been observed so far. In that setting, the case with $S=1$ is simply the classical best-choice problem. \citet[\S 2d]{GilbertMosteller1966} studied the case with $S=2$. \cite{QuineLaw1996} outlined the solution for the general problem, with a focus on the special cases $S=\{1,2,3\}$.

The optimal selection strategy of the decision-maker who is interested by ranks is defined by a series of stages, stopping points or thresholds. With $S=3$, for instance, the optimal selection strategy of the decision-maker who has access to a large pool of applicants consists of waiting until approximately $c_1=33.67\%$ of all applicants have been sampled, and then select any applicant that is the best observed so far (relative rank 1). Once $c_2=58.68\%$ of the applicants have been sampled, the decision-maker should accept the next applicant of relative rank 1 or 2. After $c_3=77.46\%$ have been passed over, the decision-maker should accept the next candidate of relative rank 1, 2, or 3.

\subsection{No Information, Minimum Expected Rank}

The related problem of minimizing the expected rank of the selected applicant was introduced by \cite{Lindley1961}. When the decision-maker can only observe the rank of each applicant relative to those who have been observed so far and the pool of applicants becomes asymptotically large, \cite{Chowetal1964} showed that the expected overall rank of the selected applicant will tend towards:
\begin{equation}\label{eqLindley1961V}
    V_{\infty} = \prod_{1}^{\infty} \left(1+\frac{2}{j} \right)^{1/(j+1)} = 3.869519 \dots
\end{equation}  %% NProduct[(1 + 2/j)^(1 + j)^(-1), {j, 1, Infinity}, WorkingPrecision -> 100] = 3.86951924139799949569416727159287663356820910607936997403285698041911350870958341075999499212362152

Let $c(x)$ denote the proportion of applicants who should be interviewed asymptotically before an applicant of relative rank $x$ or better is deemed to be a minimally acceptable (optimal) choice. As in the Gusein-Zade Problem, the optimal selection strategy of the decision-maker is characterized by a series of stages, stopping points or thresholds defined by:
\begin{equation}
    c(x) = \frac{1}{V_{\infty}} \prod_{1}^{x-1} \left(1+\frac{2}{j} \right)^{1/(j+1)},
\end{equation}
which yields $c(1)=0.25843 \dots$, $c(2)=0.44761 \dots$, $c(3)=0.56395 \dots$, $c(4)=0.64078 \dots$, $c(5)=0.69490 \dots$, $c(6)=0.73499 \dots$, and so on. In other words, the optimal selection strategy of the decision-maker consists of waiting until approximately $25.84\%$ of all applicants have been sampled, and then select any applicant that is the best observed so far (relative rank 1). Once $44.76\%$ of the applicants have been sampled, the decision-maker should accept the next applicant of relative rank 1 or 2. Once almost $56.40\%$ have been passed over, the decision-maker should accept the next applicant of relative rank 1, 2, or 3. And so on.

\subsection{Full Information, Minimum Expected Rank}

The full-information variant of the expected-rank problem is known as Robbins' problem.\footnote{Professor Herbert Robbins offered the problem at the \emph{AMS/IMS/SIAM Conference on Sequential Search and Selection in Real Time} held in Amherst, MA on June 21–27, 1990.} It is presented here for symmetry but it is out of scope for the purpose of this paper because it has not yet been solved in its general form.

In keeping with the notation used for the no-information setting, let $V(N)$ be the optimal expected overall rank of the selected applicant when there are $N$ of them. Of course, $V(1)=1$ is trivial. Similarly, $V(2)=5/4=1.25$ remains quite tractable. However, while \cite{AssafSamuelCahn1996} computed $V(3) = 1.39155 \dots$, \cite{DendievelSwan2016} derived explicitly the optimal policy leading to $V(4) = 1.49329 \dots$, and \cite{BrussFerguson1993} offered $V(5) \approx 1.5710$, the computational challenges become enormous for larger $N$ and only some bounds are known for the limiting value of the expected rank as $N$ becomes asymptotically large \citep{Bruss2005}. The general solution to the problem is therefore considered unknown.
%% V(3) = 341/144 - (13/48)*sqrt(13) = 1.39155
%% V(4) = 1.4932982948222566... (https://sites.google.com/site/yvikswan/computations-for-ds16)

In terms of lower bound, the best expected outcome a fully-informed decision-maker can hope for is to obtain the best choice (rank 1) with the probability given by Eq.~\eqref{eqPrSuccessGM1966} and the second-best choice (rank 2) the rest of the time. The lowest possible expected rank is therefore $V_{\infty} \geq 0.580164+2(1-0.580164) \approx 1.4198 \dots$. In fact, \cite{BrussFerguson1993} extrapolated that even a truncated loss function that caps the maximum punishment for any rank above $m>5$ would be expected to result at best in $V_{\infty}>1.908$. This remains the tightest available lower bound yet.

In terms of the worst case, it is obvious that the fully informed decision-maker can do no worse than Eq.~\eqref{eqLindley1961V}. In order to do better, it is useful to turn towards the class of memoryless threshold rules that only depend on the value observed at each step and the remaining number of observations. Within that class of solutions, the simple decision rule from the full-information setting of \cite{Moser1956} is an obvious candidate. Asymptotically, the expected rank within the \cite{Moser1956} setting converges from below to the upper bound $V_{\infty} < 2+\frac{1}{3}$, as demonstrated by \cite{BrussFerguson1993}. Using more complex (flexible) threshold rules, \cite{AssafSamuelCahn1996} subsequently were able to refine this upper bound to $V_{\infty}<2.3267 \dots$ but the general consensus is that significant further improvements are unlikely using memoryless threshold rules.
%% A significant part of the expected loss in terms of average rank ($\approx \frac{1}{3}$) is due to the fact that the decision rule in the \cite{Moser1956} setting often results in a choice that is not even the best so far (relative rank worst than 1). It is likely that the decision rule for Robbins' problem will reduce such occurrences.

\subsection{Summary}

Table~\ref{Tab2a} summarizes the seven prototypical decision-making settings previously mentioned. The last column refers to the numbering used by \cite{Bruss2005}. \cite{Freeman1983} and \cite{Ferguson1989} both discuss several extensions and provide ample historical context. Formal mathematical proofs are carefully explained in the electronic textbook by \cite{Ferguson2008}, while several key results are illustrated in a wonderfully intuitive manner by \citet[\S 1]{ChristianGriffiths2016}.

\begin{table}[!htb]
\centering
\begin{tabular}{l l l l}
 \hline
 Setting & Info. & Payoffs & No.\\
 \hline  \hline
 \cite{Moser1956} & Full Information & Cardinal & \\
 \cite{Bearden2006} & Best So Far & Cardinal & \\
 \cite{Mosteller1987} & Best So Far & Best Choice $\{0,1\}$ & I\\
%% \cite{Sakaguchi1984} & Best So Far & Best Choice $\{-1,0,1\}$ & & \\
 \citetalias[\S 3]{GilbertMosteller1966} & Full Information & Best Choice $\{0,1\}$ & II\\
%% \citetalias{PresmanSonin1972} & Full Information & Best Choice $\{0,1\}$ & $N \sim U(1,b)$ & \\
 \cite{GuseinZade1966} & Relative Rank So Far & Best $S$ or Better $\{0,1\}$ & \\
 \cite{Lindley1961} & Relative Rank So Far & (Expected) Rank & III\\
 \cite{Bruss2005} & Full Information & (Expected) Rank & IV\\
 \hline
\end{tabular}\\
\caption{Basic variants of optimal stopping problems.}
\label{Tab2a}
\end{table}

\section{No-Information Setting with Single Threshold}\label{Sec3}

We first turn to the no-information settings of \cite{Bearden2006} and \cite{Mosteller1987}. These are models in which the decision to select an applicant must be based only on the relative ranks of those observed.

No-information optimal stopping problems are united by the fact that their solution involves a sampling process with two or more phases delineated by a threshold rule. In the initial phase, the decision-maker sets his or her aspiration level by sequentially sampling some available applicants. The duration of this stage is determined by the optimal stopping threshold (or thresholds), a quantity that is widely studied because it defines the solution to the sequential decision problem. The subsequent phase is not studied as widely but it is equally important because this is when the decision-maker looks at additional alternatives and actively searches for a candidate that exceeds the aspiration level set during the initial phase.

\subsection{Mean and Median Number of Interviews}

Importantly, the number of applicants expected to be observed (``interviewed'') by the decision-maker before a candidate is ultimately chosen is much greater than $\sqrt{N}-1$ in the \cite{Bearden2006} setting or $N/\mathrm{e}$ (asymptotically) in the context of the Sultan's Dowry Problem \citep{Mosteller1987}. In fact, in repeated samples, the decision-maker would be expected to interview at least twice as many applicants half the time.

This can be seen by recognizing that the theoretical median of $T$ in the no-information setting, based on the probability distribution function in Eq.~\eqref{eqProbDistT}, is generally the smallest integer $T$ such that:
\begin{equation*}
\sum_{x=c^*}^{T} \frac{c^*-1}{x(x-1)} = 1 - \frac{c^*-1}{T} \geq 0.5,
\end{equation*}
which yields explicitly:
\begin{equation}\label{eqMedianNoInfo}
    \widetilde{T} = 2 \left( c^*-1 \right).
\end{equation}  %% Wolfram Alpha: sum (sqrt(N)-1)/(x*(x-1)) from x=sqrt(N) to M = 0.5 >>> M = 2*(sqrt(N)-1).

In the \cite{Bearden2006} setting, where $c^*=\sqrt{N}$, this means the median number of interviews is $\widetilde{T}= 2( \sqrt{N}-1 )$. In the Sultan's Dowry Problem of \cite{Mosteller1987}, where it is well known that the optimal threshold tends asymptotically to $N/\mathrm{e}$ \citep{Ferguson1989}, the median number of interviews will also converge to twice the threshold value: $\widetilde{T} \sim 2N/\mathrm{e}$.

The opportunity to compute more quantiles presents itself at once. More percentiles are summarized in Table~\ref{Tab1}. These percentiles apply to the entire class of no-information problems for which the optimal choice can be described in the form of an optimal threshold rule, and this threshold rule consists of skipping the first $c^*-1$ applicants before selecting the next candidate (typically a relatively best one). They apply not only asymptotically but also for smaller pools of applicants.

\begin{table}[!htb]
\centering
\begin{tabular}{r r}
 \hline
 Percentile & Interviews\\
 $\Pr(T<x)$ & $\times ( c^*-1 )$\\
 \hline  \hline
 %% $0.10$ & $1.1\overline{1}$ \\
 $0.20$ & $1.25$ \\
 Q1 = $0.25$ & $1.3\overline{3}$ \\
 $0.3\overline{3}$ & $1.5$ \\
 Median = $0.50$ & $2$ \\
 $0.6\overline{6}$ & $3$ \\
 Q3 = $0.75$ & $4$ \\
 $0.80$ & $5$ \\
%% $0.90$ & $10$ \\
 \hline
\end{tabular}
\caption{Number of interviews by percentile in the no-information setting, as a multiple of the optimal threshold value.}
\label{Tab1}
\end{table}

In the same vein, the expected number of interviewed applicants in the no-information setting is:
\begin{align}\label{eqExpectedNumberInterviews}
	\operatorname{\mathbf{E}}(T)
	& = \sum_{x=c^*}^{N-1} \frac{c^*-1}{x-1} + \frac{N(c^*-1)}{N-1} \nonumber \\
	& = (c^*-1) \left[ \frac{N}{N-1} + \sum_{x=c^*}^{N-1} \frac{1}{x-1} \right] \nonumber \\
	%% http://functions.wolfram.com/GammaBetaErf/PolyGamma2/17/01/02/0001/
	%% m=N, z=-1
	& = (c^*-1) \left[ \frac{N}{N-1} + \psi(N-1) - \psi(c^*-1) \right]
\end{align}
where $\psi(\cdot)$ is the digamma function \citep{Stern1847}.
%% Eq.~\eqref{eqExpectedNumberInterviews} is derived by leveraging the recurrence identities for distant neighbors of the digamma function:
%% \begin{align}
%%     \psi(N-1) & = \sum_{x=0}^{N-1} \frac{1}{x-1} + \psi(-1)\\
%%     \psi(c^*-1) & = \sum_{x=0}^{c^*-1} \frac{1}{x-1} + \psi(-1),
%% \end{align}
%% along with the fact that:
%% \begin{equation}
%%     \sum_{x=c^*}^{N-1} \frac{1}{x-1} = \sum_{x=0}^{N-1} \frac{1}{x-1} - \sum_{x=0}^{c^*-1} \frac{1}{x-1}.
%% \end{equation}

Eq.~\eqref{eqExpectedNumberInterviews} gives what \cite{SteinSealeRapoport2003} referred to as the expected duration of the process in terms of the total number of applicants sampled (or drawn). While they correctly recognized that it applied to all best-choice problems solved using a threshold rule, they would not yet have been aware of its applicability in other no-information settings such as the \cite{Bearden2006} setting with cardinal payoffs. In fact, Eq.~\eqref{eqExpectedNumberInterviews} applies to the entire class of no-information problems for which the optimal stopping strategy can be described in the form of an optimal threshold rule, and this threshold rule consists of skipping the first $c^*-1$ applicants before selecting the next relatively best candidate. It also applies not only asymptotically but also for smaller pools of applicants.

\subsection{Cardinal Payoffs}

Specifically in the \cite{Bearden2006} setting with cardinal payoffs, Eq.~\eqref{eqExpectedNumberInterviews} reduces to:
\begin{align}\label{eqExpectedNumberInterviewsBearden2006}
    \operatorname{\mathbf{E}}(T)
    & = (\sqrt{N}-1) \left[ \frac{N}{N-1} + \psi(N-1) - \psi(\sqrt{N}-1) \right] \nonumber \\
    & \approx (\sqrt{N}-1) \left[ 1 + 0.5 \ln{N} \right] + 1.5,
\end{align}
where the approximation reflects the second-order generalized Puiseux series expansion \citep{Puiseux1850} and is always accurate up to the nearest integer.
%% Wolfram Alpha: series (sqrt(X)-1)*( X/(X-1) + Digamma(X-1) - Digamma(sqrt(X)-1) )

As a side note, it should be obvious that the proportion of interviewed applicants tends to zero asymptotically in the \cite{Bearden2006} setting since both the median and the mean only grow as the square root of $N$.
%% in the sense that:
%% \begin{equation}\label{eqMeanBearden2006}
%% 	\lim_{N \to \infty} \frac{\operatorname{\mathbf{E}}(T)}{N} \approx \lim_{N \to \infty} \frac{( \sqrt{N}-1 ) \left[ 1 + 0.5 \ln{N} \right]}{N} + \frac{1.5}{N} = 0
%% %% \end{equation}
%% and
%% \begin{equation}\label{eqMedianBearden2006}
%% 	\lim_{N \to \infty} \frac{\widetilde{T}}{N} = \lim_{N \to \infty} \frac{2(\sqrt{N}-1)}{N} = 0.
%% \end{equation}

\subsection{Large Number of Applicants}

When the number of applicants is large and the threshold rule in the no-information setting dictates that a fixed proportion $c^*/N = x$ of these applicants will be passed over before a candidate is finally chosen, Eq.~\eqref{eqExpectedNumberInterviews} simplifies asymptotically to:
\begin{equation}\label{eqExpectedAsympt}
    \frac{\operatorname{\mathbf{E}}(T)}{N} \sim x \left[ 1 + \ln \left( \frac{1}{x} \right) \right].
\end{equation}

Eq.~\eqref{eqExpectedAsympt} illustrates well the two-phase process that is involved with no-information optimal stopping problems characterized by a threshold rule \citep{Todd2007}. By definition, the duration of the initial stage is the optimal stopping threshold ($x$). Asymptotically, the expected duration of the second stage is proportional to the first stage: it becomes $\ln{(1/x)}$ as long.

An inquisitive observer might wonder in what circumstances roughly half of all applicants would be expected to be interviewed in the no-information setting, i.e. $\mathbf{E}(T)/N=p=0.5$. If we use $\mathbf{W}_{\scriptscriptstyle{-1}}(\cdot)$ to denote the lower branch of the Lambert~W function defined implicitly by $z=\mathbf{W}(z\mathrm{e}^z)$ \citep{Corlessetal1996} and define a function $w(p) = 1+\mathbf{W}_{\scriptscriptstyle{-1}}(-p/\mathrm{e})$, it is easy to verify that an optimal threshold of $x = \mathrm{e}^{w(0.5)} \approx 0.186682$ would do the trick. More generally, a threshold of $\mathrm{e}^{w(p)}$ will result in a proportion $p$ of all applicants expected to be interviewed on average, asymptotically.

\subsubsection{Sultan's Dowry Problem}

In the context of the classical Secretary Problem or the Sultan's Dowry Problem of \cite{Mosteller1987}, \cite{QuineLaw1996}, \cite{SteinSealeRapoport2003}, and \cite{MazalovPeshkov2004} all previously reported that the expected number of interviews converges asymptotically to $2N /\mathrm{e} \approx 0.736N$, which is also the median. This result can be obtained directly using Eq.~\eqref{eqExpectedAsympt}. Taking the well-established asymptotic threshold of $x = 1/\mathrm{e}$ as given, we have:
\begin{align}\label{eqMeanMosteller1987}
	\lim_{N \to \infty} \frac{\operatorname{\mathbf{E}}(T)}{N}
%%	& = \lim_{N \to \infty} \frac{(c^*-1)}{N} \left[ \frac{N}{N-1} + \psi(N-1) - \psi(c^*-1) \right] \nonumber \\
	& = \left( \frac{1}{\mathrm{e}} \right) \left[ 1 + \ln (\mathrm{e}) \right]
	= \frac{2}{\mathrm{e}} \approx 0.7358.
\end{align}  %% Wolfram Alpha: limit digamma(N-1)-digamma(N/e) as N->infinity = 1

On average, if the Sultan has a large number of daughters, the commoner will see approximately 73.6\% of them by the time he asks one in marriage. Half the time, he will see more. Half the time, he will see fewer.

\begin{figure}
\centering
\includegraphics{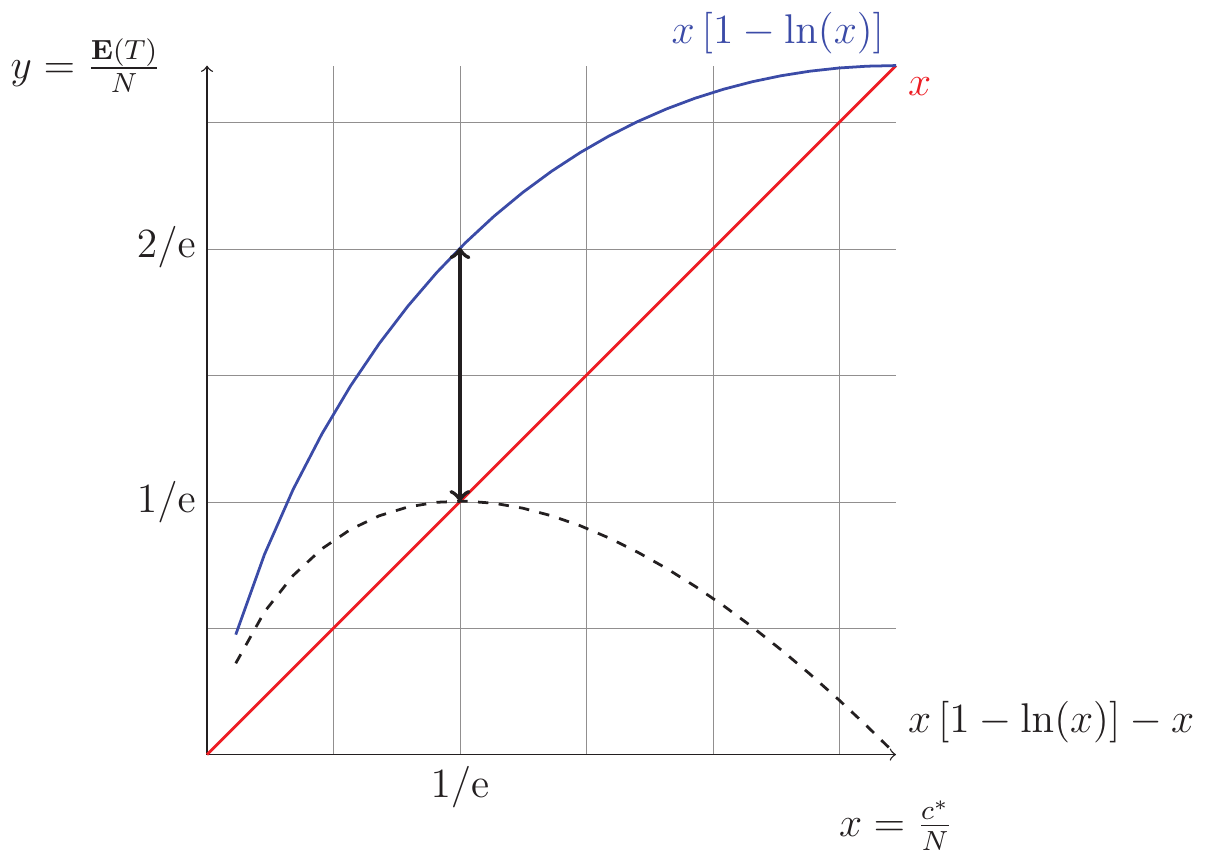}
%% \begin{tikzpicture}[domain=0:1,scale=7,font=\large]
%%     \draw[step=0.1835,ultra thin,color=gray] (0,0) grid (1,1);
%%     \draw[->] (0,0) -- (1,0);
%%     \draw[->] (0,0) -- (0,1);
%%     \draw[thick,color=red] plot[id=x] function{x}
%%         node[below right] {$x$};
%%     \draw[thick,color=blue] plot[id=Ex] function{x*(1-log(x))}
%%         node[above left] {$x \left[ 1 - \ln(x) \right]$};
%%     \node[below] at (0.367,0) {$1/\mathrm{e}$};
%%     \node[left] at (0,0.367) {$1/\mathrm{e}$};
%%     \node[left] at (-0.05,1) {$y = \frac{\operatorname{\mathbf{E}}(T)}{N}$};
%%     \node[below] at (1,-0.05) {$x = \frac{c^*}{N}$};
%%     \node[left] at (0,0.736) {$2/\mathrm{e}$};
%%     \draw[very thick,<->] (0.367,0.367) -- (0.367,0.736);
%%     \draw[dashed,thick,color=black] plot[id=Distance] function{x*(1-log(x))-x}
%%         node[above right] {$x \left[ 1 - \ln(x) \right] - x$};
%% \end{tikzpicture}
\caption{Maximum distance between the optimal threshold ($c^*/N = x$) and the expected proportion of interviewed applicants as given by Eq.~\eqref{eqExpectedAsympt} is achieved asymptotically when $x=\mathrm{e}^{-1}$. This is precisely the solution of the classical Secretary Problem.}
\label{Fig1}
\end{figure}

One element that separates the classical Secretary Problem from all other optimal stopping problems that are solved using a threshold rule, is that it is the setting in which the decision-maker can expect to interview the most applicants \emph{after} the optimal threshold has been reached. This is illustrated by Figure~\ref{Fig1}.
%% It is also the only setting where the (second) selection stage is expected to be as long as the initial sampling stage.

\subsubsection{Dowry Problem With Multiple Picks}

Even if the decision-maker can make two or more picks in the \cite{Mosteller1987} setting with the hope that one of them will be the single best overall applicant, the optimal threshold for the last choice asymptotically remains the same as in the single-choice setting, as shown in \citet[\S 2b]{GilbertMosteller1966}. The expected number of interviews therefore remains the same and Eq.~\eqref{eqMeanMosteller1987} continues to apply asymptotically.

\subsubsection{Win-Lose-Draw Marriage Problem}

If we were to consider the marriage problem of \cite{Sakaguchi1984}, where the payoff is $1$ if the decision-maker marries the best, $-1$ if the decision-maker marries any other candidate who is not the best, and the decision-maker can settle for a neutral payoff of $0$ by not getting married, the optimal threshold for an asymptotically large pool size would be instead $x = 1/\sqrt{\mathrm{e}} \approx 0.607$ \citep{FergusonRejoinder1989}.

In that case, the proportion of applicants who can expect to be interviewed converges asymptotically to:
\begin{equation*}\label{eqMeanSakaguchi1984}
	\lim_{N \to \infty} \frac{\operatorname{\mathbf{E}}(T)}{N}
%%	& = \lim_{N \to \infty}\ \frac{(c^*-1)}{N} \left[ \frac{N}{N-1} + \psi(N-1) - \psi(c^*-1) \right] \nonumber \\
	= \left( \frac{1}{\sqrt{\mathrm{e}}} \right) \left[ 1 + \ln(\sqrt{\mathrm{e}}) \right]
	= \frac{3}{2 \sqrt{\mathrm{e}}} \approx 0.9098
\end{equation*}  %% Wolfram: limit digamma(N-1)-digamma(N/sqrt(e)) as N->infinity = 0.5

In other words, a whopping 91.0\% of all applicants can expect to be interviewed in the \cite{Sakaguchi1984} setting, on average. Of course, it makes sense that most applicants can expect to be interviewed because the decision-maker in that setting will actually end up interviewing all $N$ applicants without choosing one with probability approaching $1/\sqrt{\mathrm{e}} \approx 0.607$.

Conditional on an applicant being chosen in the first place, however, the average proportion of applicants interviewed by the decision-maker in the \cite{Sakaguchi1984} setting will be approximately:
\begin{equation*}
    \frac{\operatorname{\mathbf{E}}(T \, | \, \text{Get Married})}{N} \sim  \frac{1}{2(\sqrt{\mathrm{e}}-1)} \approx 0.7707,
\end{equation*}
which remains close to $2/\mathrm{e} \approx 0.7358$, the proportion of applicants expected to be interviewed in the classical Secretary Problem.

\subsubsection{Postdoc Problem}

In the Postdoc Problem, the decision-maker tries to choose the second-best applicant instead of the best overall because, as the story goes, it is expected that ``the best applicant will receive and accept an offer from Harvard'' \citep{Vanderbei1995}. A solution specific to this second-best problem was presented by \cite{Rose1982a} and the more general problem of selecting the $a\textsuperscript{th}$ best was further studied by \cite{Szajowski1982}, \cite{Rose1982b} and \cite{Linetal2019}, among others.

Given an arbitrary threshold rule where the first $c$ applicants are sampled but never selected, the probability of successfully choosing the second-best applicant overall out of $N$ applicants is:
\begin{equation}
    \Pr{(\text{Success})} = \frac{c}{N} \sum_{k=c+1}^{N} \frac{(c-1)}{(k-1)(k-2)} = \frac{c(N-c)}{N(N-1)}
\end{equation}

Maximizing this probability comes down to maximizing the numerator $c(N-c)$, which is achieved by setting $c^*=N/2$ when $N$ is even. In other words, the optimal strategy for the decision-maker in the Postdoc Problem consists of sampling the first $c^*= \lfloor N/2 \rfloor$ applicants before accepting the next relatively second-best candidate who comes along. As implied by this solution, it is never strictly preferable to select a relatively best candidate instead of waiting for a relatively second-best one \citep{Bayonetal2018}. The probability of success with this strategy asymptoticaly tends to $\Pr{(c^*)} \sim 1/4$.

Given this threshold rule, the probability that the $x\textsuperscript{th}$ applicant will be relatively second-best and will be selected in the Postdoc Problem is:
\begin{equation}\label{eqPrXPostdoc}
    \Pr(T=x) = \frac{c^*-1}{x(x-1)} = \frac{N-1}{2x(x-1)} \quad \text{for }\ \frac{N}{2}<x<N \\
\end{equation}

The probability that the decision-maker in the Postdoc Problem will fail to find a relatively second-best candidate even after sampling the first $N-1$ applicants is:
\begin{equation}\label{eqPrNPostdoc}
    \Pr(T=N) = 1 - \sum_{x=c^*+1}^{N-1} \frac{N-1}{2x(x-1)} = 1 - \frac{1}{2} + \frac{1}{N} = \frac{1}{2} + \frac{1}{N}.
\end{equation}

It directly follows that the median is $N$: all Postdoc applicants will be interviewed half the time.

The expected number of interviewed applicants is obtained by combining Eq.~\eqref{eqPrXPostdoc} and Eq.~\eqref{eqPrNPostdoc}:
\begin{align*}
    \operatorname{\mathbf{E}}(T)
    & = N \cdot \Pr(T=N) + \sum_{x=c^*+1}^{N-1} x \cdot \Pr(T=x) \nonumber\\
%%    & = \frac{N}{2} + \frac{N}{2(N-1)} + \frac{N}{2} \left\{ \psi{(N-1)} - \psi{\left( \frac{N}{2} \right)} \right\} \nonumber\\
%%    & = \frac{N}{2} \left\{ 1 + \frac{1}{N-1} + \psi{(N-1)} - \psi{\left( \frac{N}{2} \right)} \right\} \nonumber\\
    & = \frac{N-1}{2} \left\{ \frac{N}{N-1} + \psi{(N-1)} - \psi{\left( \frac{N}{2} \right)} \right\} + 1 \nonumber\\
    & \sim \frac{N}{2} \left\{ 1 + \ln(2) \right\} + 1
\end{align*}

In other words, the decision-maker who is looking for the second-best overall is expected to interview approximately $\{ 1+\ln(2) \}/2 \approx 84.66\%$ of all applicants on average. Of course, this asymptotic result can obtained directly using Eq.~\eqref{eqExpectedNumberInterviews} and Eq.~\eqref{eqExpectedAsympt} with $c^*/N=1/2$.

%% Suppose that the $k\textsuperscript{th}$ applicant is ranked second-best among the $k$ applicants seen so far, and is therefore a viable candidate. The probability that this candidate will remain second-best after all $N$ applicants are interviewed is $\Pr(k \, | \, \text{Relatively Second-Best}) = k(k-1)/(N^2-N)$. If the $k\textsuperscript{th}$ applicant is instead relatively best, it may still drop to second place if there is a better applicant among those who have yet to be interviewed. This will occur with probability $\Pr(k \, | \, \text{Relatively Best}) = k(N-k)/(N^2-N)$. The probability of success is higher with the second-best as long as $k>c^*=(N+1)/2$, which is satisfied by $k>N/2$ as long as $N$ is even. After $(N+1)/2$ of all applicants have been interviewed, the decision-maker would therefore prefer a relatively second-best candidate to a relatively best one.

\subsubsection{Random Number of Applicants}

Suppose that the actual number of applicants ($N$) was not precisely known in the Secretary or Sultan's Dowry Problem but was distributed instead uniformly on $\{1,\dots,b\}$ with a known maximum pool size $b$. (Of note, the actual number of applicants is expected to be $\operatorname{\mathbf{E}}(N) = b/2$ on average.) The decision-maker wins by selecting the very best applicant. Otherwise, if the decision-maker passes up on the last applicant without making a selection, he or she loses. This is the Presman-Sonin best-choice problem \citep{PresmanSonin1972}.

Asymptotically, when $b$ is large, the threshold as a proportion of the maximum pool size converges to $c^*/b = \mathrm{e}^{-2} \approx 0.1353$ or, as a proportion of the expected pool size, $\operatorname{\mathbf{E}}(c^*/N) = 2\mathrm{e}^{-2} \approx 0.2707$. Incidentally, the latter is also the average probability of successfully selecting the single best applicant \citep{Freeman1983}.

Let $n$ denote the realized value of the random variable $N$. Since $N$ is uniformly distributed over $\{1,\dots,b\}$, each value is equally likely with $\Pr{(N=n)}=1/b$. In order to evaluate the mean proportion of applicants who can expect to be interviewed out of the $n$ actual applicants (not the theoretical upper bound $b$) in this particular variant of the Secretary Problem, we have to consider two possibilities depending on the actual realization of $N$. When $n \leq b/\mathrm{e}^2$, the interview process ends prematurely because all $n$ applicants are interviewed during the sampling phase, before the optimal stopping threshold is even reached. This eventuality is expected to occur with probability $1/\mathrm{e}^2$. When $n > b/\mathrm{e}^2$, the same reasoning that led to Eq.~\eqref{eqExpectedNumberInterviews} continues to apply so we have:
\begin{align*} %% \label{eqExpectedPresmanSonin}
	\operatorname{\mathbf{E}}(T \, | \, N=n)
	& = \sum_{k=b/\mathrm{e}^2+1}^{n} \frac{b}{\mathrm{e}^2(k-1)} + \frac{n}{\mathrm{e}^2} \nonumber\\
	& = \frac{b}{\mathrm{e}^2} \left[ \psi{(n)} - \psi{(b/\mathrm{e}^2)} + \frac{n}{b} \right] \nonumber\\
	& \sim \frac{b}{\mathrm{e}^2} \left[ 2 + \ln{\left( \frac{n}{b} \right)} + \frac{n}{b} \right]
\end{align*}

After dividing by $n$, passing to the limit, evaluating the integral from $n/b=1/\mathrm{e}^2$ to 1, and adding $1/\mathrm{e}^2$ to reflect the $n \leq b/\mathrm{e}^2$ case, we obtain the average proportion of applicants who can expect to be interviewed in the \cite{PresmanSonin1972} setting:
\begin{align}\label{eqMeanPresmanSonin1972}
    \lim_{N \to \infty} \frac{\operatorname{\mathbf{E}}(T)}{n}
	& = \frac{1}{\mathrm{e}^2} + \int_{\mathrm{e}^{-2}}^{1} \frac{2 + \ln(x) + x}{x\mathrm{e}^2} dx \nonumber\\
	& = \frac{1}{\mathrm{e}^2} + \frac{3\mathrm{e}^2 - 1}{\mathrm{e}^4} \nonumber\\
	& = \frac{4\mathrm{e}^2-1}{\mathrm{e}^4} \approx 0.5230.  %% 0.52302549405
\end{align}

It comes somewhat as a surprise that Eq.~\eqref{eqMeanPresmanSonin1972} coincides with the \emph{median} proportion of applicants expected to be interviewed in the presence of a known number of applicants and an optimal threshold rule with $c^*/N=2/e^2$, up to a difference of only $\mathrm{e}^{-4} \approx 0.0183$.

The challenge of evaluating the median in the context of the Presman-Sonin best-choice problem with $N \sim U\{1,\dots,b\}$ is left as an open problem.

\subsubsection{Uncertain Employment}

Another way to extend the classical Secretary Problem is to allow for the possibility that the chosen candidate may be unavailable or may refuse an offer of employment with a fixed probability. This is the framework introduced by \cite{Smith1975}.

Within the context of the Sultan's Dowry Problem, suppose that any of the $N$ daughters can turn down the commoner and refuse his marriage offer with probability $1-p$. When this occurs, the search for a consenting bride continues. It turns out that the optimal threshold asymptotically tends to $c^* \sim Np^{1/(1-p)}$ \citep{Smith1975}, in the sense that:
\begin{equation}\label{eqSmith1975Threshold}
    \lim_{N \to \infty} \frac{c^*}{N} = p^{1/(1-p)}
\end{equation}
As a side note, Eq.~\eqref{eqSmith1975Threshold} is also the probability of marrying the best daughter using the optimal threshold rule.

In that case, the proportion of applicants who can expect to be interviewed converges asymptotically to:
\begin{align*} %% \label{eqMeanSmith1975}
	\lim_{N \to \infty} \frac{\operatorname{\mathbf{E}}(T)}{N}
%%	& = \lim_{N \to \infty}\ \frac{c^*}{N} \left[ \frac{N}{N-1} + \psi(N-1) - \psi(c^*) \right] \nonumber \\
	& = p^{1/(1-p)} \left[ 1 - \ln \left( p^{1/(1-p)} \right) \right]
	= p^{1/(1-p)} \left[ 1-\frac{\ln(p)}{1-p} \right]
\end{align*}  %% Wolfram: limit p^(1/(1-p)) * (1-log(p)/(1-p)) as p -> 1 = 2/e

When $p=0.5$, for instance, $c^*/N = 1/4$ and $\operatorname{\mathbf{E}}(T)/N = [1+2\ln{(2)}]/4 = 0.596574$. As $p \to 1$, we retrieve as expected the solution for the classical Secretary Problem: $c^*/N \sim \mathrm{e}^{-1}$ and $\operatorname{\mathbf{E}}(T)/N \sim 2/\mathrm{e}$.

\subsubsection{No-Information Duration Problem}

\cite{FergusonHardwickTamaki1992} introduced the duration problem, where the decision-maker must pick one and only one candidate and the payoff is proportional to the length of time the chosen candidate remains relatively best. In the no-information setting, the decision-maker can only observe whether an applicant is best so far. Like in other best-choice settings, it only makes sense for the decision-maker to select a relatively best applicant. The decision-maker will then receive an additional payoff of one for each applicant interviewed subsequently, until another relatively best candidate comes up (if any).

In the simplest form of the duration problem, the selected candidate does not necessarily need to be the best overall. Asymptotically, the optimal stopping threshold in this no-information duration problem is $c^*/N = e^{-2} \approx 0.1353$ \citep{FergusonHardwickTamaki1992}. In other words, the decision-maker should wait until 13.53\% of all applicants have been sampled and select the next relatively best candidate that becomes available. While the threshold rule in the duration problem is reminiscent of the optimal threshold rule in the Presman-Sonin best-choice problem where $N \sim U(1,b)$ \citep{PresmanSonin1972}, the expected proportion of interviewed applicants is more straightforward. It tends to:
\begin{align*}\label{eqMeanDurationProblem1992A}
	\lim_{N \to \infty} \frac{\operatorname{\mathbf{E}}(T)}{N}
	& = \frac{1}{\mathrm{e}^{2}} \left[ 1 + \ln \left( \mathrm{e}^{2} \right) \right]
	= \frac{3}{\mathrm{e}^2} \approx 0.4060. %% 0.40600584971
\end{align*}

We note in passing that the median is simply $\widetilde{T}/N = 2e^{-2} = 0.2707$, which is also the expected payoff for the decision-maker asymptotically \citep{FergusonHardwickTamaki1992}.

\subsubsection{Best-Choice Duration Problem}

One variant of the duration problem also discussed by \cite{FergusonHardwickTamaki1992} makes the decision-maker's payoff contingent on the selected candidate turning out to be the best overall. This is the best-choice duration problem. In that case, the payoff of selecting the $k\textsuperscript{th}$ applicant is $(N-k+1)/N$ if it is the best overall, and zero otherwise. In other words, the reward depends on the proportion of time the decision-maker is in possession of the best overall applicant.

The asymptotically optimal stopping threshold in the best-choice duration problem is the solution to $2x-\ln(x)=2$ \citep{FergusonHardwickTamaki1992}. If we use $\mathbf{W}(\cdot)$ to denote the upper (principal) branch of the Lambert~W function defined implicitly by $z=\mathbf{W}(z\mathrm{e}^z)$ \citep{Corlessetal1996,Wolfram}, this solution can be expressed as $c^*/N = - \frac{1}{2}\mathbf{W}(-2 / \mathrm{e}^{2}) = 0.20318786 \dots$.\footnote{As noted by \cite{Bayonetal2018}, the same constant $0.20318 \dots$ appears in the unrelated Daley-Kendall model \citep{DaleyKendall1965}, where it is known as the rumour's constant. Its decimal expansion is given by the OEIS sequence \href{https://oeis.org/A106533}{A106533}.} In other words, the decision-maker should wait until 20.32\% of all applicants have been sampled and select the next relatively best candidate that becomes available. %% 0.20318787

The expected proportion of interviewed applicants, in that case, asymptotically tends to:
\begin{equation*} %% \label{eqMeanDurationProblem1992B}
	\lim_{N \to \infty} \frac{\operatorname{\mathbf{E}}(T)}{N}
%%	& = - \frac{1}{2} \mathbf{W}(-2 / \mathrm{e}^{2}) \left[ 1 + \ln \left( - \frac{1}{2} \mathbf{W}(-2 / \mathrm{e}^{2}) \right) \right] \nonumber\\
	= - \frac{1}{2} \mathbf{W}(-2 / \mathrm{e}^{2}) \left[ 3 + \mathbf{W}(-2 / \mathrm{e}^{2}) \right] \nonumber\\
%%	& = -0.5 \left[ 3\cdot \mathbf{W}(-2 / \mathrm{e}^{2}) + {\mathbf{W}(-2 / \mathrm{e}^{2})}^2 \right] \nonumber\\
	\approx 0.5270.  %% 0.52699319612
\end{equation*}

For its part, the median is twice the optimal threshold: $\widetilde{T}/N = -\mathbf{W}(-2 / \mathrm{e}^{2}) \approx 0.4064$. 

It is noteworthy that the exact same optimal stopping threshold applies to the variant of the Secretary Problem where the decision-maker incurs interview costs of $1/N$ per applicant, as shown by \cite{Bayonetal2018}. For an arbitrary stopping threshold $c$, let $x$ denote the asymptotic value of $c/N$. Consequently, the expected payoff function for the decision-maker when there are interview costs of $1/N$ per applicant is:
\begin{align}
    EV(c) & = \frac{c-1}{N} \sum_{k=c}^{N} \frac{1-\frac{k}{N}}{k-1} \nonumber\\
    & = \frac{c-1}{N} \cdot \frac{(N-1) \left[ \psi(N) - \psi(c-1) \right] - (N-c+1)}{N} \nonumber\\
    & \sim x \left[ x-1 - \ln(x) \right]
\end{align}

This is maximized when $2x-\ln(x)=2$, or $x=- \frac{1}{2}\mathbf{W}(-2 / \mathrm{e}^{2}) = 0.20318786 \dots$ Therefore, the mean and median proportion of interviewed applicants in the Secretary Problem with interview costs are also as described above.

In fact, the same optimal stopping threshold also appears in the \cite{Bearden2006} setting with a decision cost proportional to $0.5(N-k+1)/(N-c+1)$ and an asymptotically large pool of applicants \citep{Szajowski2009}. More generally, when the decision cost is proportional to $C(N-k+1)/(N-c+1)$, the optimal stopping threshold is the unique solution in $(0,1)$ to $\ln(c)=b(c-1)$, where $b = 1+1/2C$. This yields explicitly $c^* = - \mathbf{W}(-b \mathrm{e}^{b}) / b$.

\section{Multi-Stopping Rank Problems}\label{Sec4}

As stated in subsection~\ref{BestS}, the optimal selection strategy of the decision-maker who is interested by (relative) ranks is defined by a series of stages $\mathbf{c}$ instead of a single threshold $c^*$. This includes, for example, the settings of \cite{GuseinZade1966} and \cite{Lindley1961}. Eq.~\eqref{eqMedianNoInfo} and Eq.~\eqref{eqExpectedNumberInterviews}, including the asymptotic version in Eq~\eqref{eqExpectedAsympt}, do not work in these cases. Fortunately, the problem of figuring out how many interviews will be conducted in such multi-stopping problems was previously solved by \cite{Yeo1997}.

\subsection{Median in Rank Problems}

Let $m = \widetilde{T}/N$ represent the median proportion of interviewed candidates. Asymptotically, we can rely on the cumulative probability distribution derived by \cite{Yeo1997} and solve:
\begin{equation}\label{eqMedianNoInfoRanks}
F(m  \, | \, \mathbf{c}) = 1 - \frac{1}{m^d} \prod_{i=1}^{d} c_i = 0.5 \; \textrm{,} \quad c_d< m <c_{d+1}
\end{equation}

When the median proportion falls between $c_2$ and $c_3$, the solution to Eq.~\eqref{eqMedianNoInfoRanks} is given by $ m = \widetilde{T}/N = \sqrt{2 c_1 c_2}$. When the median proportion falls between $c_3$ and $c_4$, the solution is instead $m = \widetilde{T}/N = ( 2 c_1 c_2 c_3)^{1/3}$. And so on. In other words, the median in rank problems such as those posed by \cite{Lindley1961} and \cite{GuseinZade1966} is simply the geometric mean of the stopping points that are passed over \emph{times} a multiplier of $2^{(1/d)}$, where $d$ is the number of stopping points that are passed over on the way to the median. In that context, Eq.~\eqref{eqMedianNoInfo} is simply a special case of Eq.~\eqref{eqMedianNoInfoRanks} with $\mathbf{c}=c^*$ (and consequently $d$=1).

Empirically, for the minimum expected rank problem of \cite{Lindley1961} and for many reasonable values of the minimum target rank $S$ in the \cite{GuseinZade1966} setting, it turns out that the median proportion of interviewed applicants typically falls after $x_2$ but before $x_4$. In other words, rank problems are typically expected to stop relatively early on average. For example, the median proportion of applicants who can expected to be interviewed in the asymptotic version of the minimum expected rank problem of \cite{Lindley1961} is approximately $m \approx \sqrt{2 \cdot 0.25843 \cdot 0.44761} = 0.48099$. In the \cite{GuseinZade1966} setting, $m \approx \sqrt{2 \cdot 0.3129 \cdot 0.4367} = 0.52276$ when $S=10$ and $m \approx ( 2 \cdot 0.3008 \cdot 0.3702 \cdot 0.4242)^{1/3} = 0.45545$ when $S=25$.

Of course, the opportunity to compute other percentiles using Eq.~\eqref{eqMedianNoInfoRanks} presents itself at once. For the $p$ percentile, the applicable multiplier would be $1/(1-p)^{(1/d)}$. The rightmost column of Table~\ref{Tab1} shows the applicable values of $1/(1-p)$ for common percentiles.

\subsection{Mean in the Gusein-Zade Setting}

The explicit formula giving the number of applicants who can expect to be interviewed in the \cite{GuseinZade1966} setting for any given target rank $S$ and an applicant pool of size $N$ was also previously derived by \cite{Yeo1997}.
%% It is:
%% \begin{align}
%%     \operatorname{\mathbf{E}}(T) = (k_1-1) &
%%     \left[
%%         \begin{array}{ll}
%%          \sum_{d=2}^{S} \frac{d}{(d-1)!} \prod_{i=2}^{d-1} (k_i-i) \sum_{i=0}^{d-2} (-1)^{d-1} {d-2 \choose i} \left( \frac{1}{k_d-i-2} - \frac{1}{k_{d+1}-i-2} \right)\\
%%         + \sum_{j=k_1-1}^{k_2-2} \frac{1}{j} + \frac{N}{N-1} \prod_{i=2}^{S} \frac{k_i-i}{N-i}
%%         \end{array}
%%       \right]
%% \end{align}

Of interest, when $N$ becomes asymptotically large, the proportion of applicants who can expect to be interviewed is:
\begin{align}\label{eqMeanYeo}
    \lim_{N \to \infty} \frac{\operatorname{\mathbf{E}}(T)}{N}
    ={}& c_1 \cdot \ln{ \left( \frac{c_2}{c_1} \right) } \nonumber\\
    & \negmedspace+ \sum_{d=2}^{S} \frac{d}{d-1} \left( \prod_{i=1}^{d} c_i \right) \left( \frac{1}{{c_d}^{d-1}} - \frac{1}{{c_{d+1}}^{d-1}} \right) \nonumber\\
    & \negmedspace+ \prod_{i=1}^{S} c_i,
\end{align}
where $\mathbf{c}=\{c_1,c_2,\dots,c_S\}$ are the optimal thresholds expressed as a proportion of all $N$ applicants.

Specifically in the \cite{GuseinZade1966} setting, the proportion of applicants who can expect to be interviewed is therefore approximately $0.6892$ in the $S=2$ case, $0.6564$ in the $S=3$ case, and $0.6102$ in the $S=5$ case \citep{Yeo1997}.\footnote{\cite{QuineLaw1996} previously reported that $\operatorname{\mathbf{E}}(T)/N$ asymptotically approaches 0.718 in the $S=2$ case and 0.713 in the $S=3$ case but there is a typo in their formulas.} The proportion shrinks down to approximately $0.5450$ in the $S=10$ case and $0.5095$ in the $S=15$ case \citep{Goldenshlugeretal2019}. By the time $S=25$, the 
\emph{ex ante} probability that any given applicant will be interviewed has shrunk down to $0.4700$. Naturally, one may wonder what happens to this proportion as the decision-maker becomes increasingly lenient and $S \to \infty$ (along with $N$). It turns out that the proportion of applicants who can expect to be interviewed tends to the limiting value $t^* \approx 0.2834$ as $S \to \infty$, an earlier result due to \cite{FrankSamuels1980}.

The same way Eq.~\eqref{eqMedianNoInfoRanks} is a generalization of the single-threshold median given by Eq.~\eqref{eqMedianNoInfo}, Eq.~\eqref{eqMeanYeo} generalizes for multi-stopping problems the single-threshold mean given be Eq.~\eqref{eqExpectedAsympt}, in the sense that we retrieve back Eq.~\eqref{eqExpectedAsympt} when $c_1=x$ and $c_2=c_3=\dots=1$ in Eq.~\eqref{eqMeanYeo}.

\subsection{Mean in Minimum Rank Problem}

Importantly, Eq.~\eqref{eqMeanYeo} applies not only to the \cite{GuseinZade1966} setting but also to any no-information setting where the threshold rule consists of a series of stopping points instead of a single cutoff point. This includes the \cite{Lindley1961} problem, where the decision-maker is trying to minimize the overall rank of the selected candidate.

%% Applying Eq.~\eqref{eqMeanYeo} to the rank problem of \cite{Lindley1961} yields:
%% \begin{equation}
%%    \lim_{N \to \infty} \frac{\operatorname{\mathbf{E}}(T)}{N} \approx 0.5065.
    %% 0.50646667...
%% \end{equation}
%% The reasoning is as follows.

Given that $c_1 = 1/V_{\infty}$, $c_2 = \sqrt{3}/V_{\infty}$ and $\prod c_i \to 0$ asymptotically in the multi-stopping minimum rank problem of \cite{Lindley1961}, Eq.~\eqref{eqMeanYeo} predicts:
\begin{align*}
   \lim_{N \to \infty} \frac{\operatorname{\mathbf{E}}(T)}{N}
   & = c_1 \ln{(\frac{c_2}{c_1})} + 2 c_1 - \frac{1}{2} \frac{c_1 c_2}{c_3} - \frac{1}{6} \frac{c_1 c_2 c_3}{{c_4}^2} - \frac{1}{12} \frac{c_1 c_2 c_3 c_4}{{c_5}^3} \dots \\
   & = \frac{1}{V_{\infty}} \left\{ 0.5 \ln{(3)} + 2 - \sum_{k=1}^{\infty} \frac{1}{k(k+1)} \prod_{j=1}^{k} {\left( 1+\frac{2}{j+1} \right)}^{\frac{-j}{j+2}} \right\} \\
   & = \frac{1}{V_{\infty}} \left\{ 0.5 \ln{(3)} + 2 - \sum_{k=1}^{\infty} \frac{1}{k(k+1)} \prod_{j=2}^{k+1} {\left( 1+\frac{2}{j} \right)}^{\frac{-j+1}{j+1}} \right\}
\end{align*}

The first few terms of the infinite sum are:
\begin{align*}
    \sum_{k=1}^{\infty} \frac{1}{k(k+1)} \prod_{j=2}^{k+1} {\left( 1+\frac{2}{j} \right)}^{\frac{-j+1}{j+1}} = & \frac{1}{2^{(4/3)}} + \frac{1}{2^{(4/3)} \sqrt{15}} + \frac{1}{2^{(26/15)} 3^{(6/10)} \sqrt{15}} \\ & + \frac{1}{2^{(26/15)} 3^{(1/10)} 5^{(5/6)} 7^{(2/3)}} \dots
%% \\ & + \frac{1}{2^{(12/5)} 3^{(4/15)} 5^{(5/6)} 7^{(2/3)}} \approx & 0.569
\end{align*}

The first 50 terms add up to approximately $0.5895$. The remaining terms, for their part, necessarily add up to less than $5/70278$ since:
\begin{align*}
    \sum_{k=51}^{\infty} \frac{1}{k(k+1)} \left[ \prod_{j=2}^{k+1} {\left( 1+\frac{2}{j} \right)}^{\frac{-j+1}{j+1}} \right]
    & = \sum_{k=51}^{\infty} \frac{1}{k(k+1)} \left[ \prod_{j=2}^{k+1} {\left( 1+\frac{2}{j} \right)}^{-1} \cdot \prod_{j=2}^{k+1} {\left( 1+\frac{2}{j} \right)}^{\frac{2}{j+1}} \right] \\
    & \leq \sum_{k=51}^{\infty} \frac{1}{k(k+1)} \left[ \frac{6}{(k+2)(k+3)} \cdot \prod_{j=2}^{\infty} {\left( 1+\frac{2}{j} \right)}^{\frac{2}{j+1}} \right] \\
    & \approx \sum_{k=51}^{\infty} \frac{6}{k(k+1)(k+2)(k+3)} \cdot 4.99106 \\
    & = \frac{4.99106}{70278} = 0.0000142 \dots
\end{align*}

Up to four decimals, the average proportion of interviewed applicants in the \cite{Lindley1961} setting is therefore:
\begin{equation*}
    \lim_{N \to \infty} \frac{\operatorname{\mathbf{E}}(T)}{N} \approx \frac{0.5 \ln{(3)} + 2 - 0.5895}{V_{\infty}}  \approx \frac{1.9598}{3.8695} = 0.5065.
\end{equation*}

%% This makes use of the fact that:
%% \begin{align*}
%%     \prod_{j=2}^{k+1} \left( 1+\frac{2}{j} \right)^{\frac{-j+1}{j+1}} & = \prod_{j=2}^{k+1} {\left( 1+\frac{2}{j} \right)}^{\frac{2}{j+1}} \cdot \prod_{j=2}^{k+1} {\left( 1+\frac{2}{j} \right)}^{-1} \\
%%     & \leq \prod_{j=2}^{\infty} {\left( 1+\frac{2}{j} \right)}^{\frac{2}{j+1}} \cdot \prod_{j=2}^{k+1} {\left( 1+\frac{2}{j} \right)}^{-1} \\
%%     & \approx 4.99106 \cdot \frac{6}{(k+2)(k+3)}
%% \end{align*}

In other words, the decision-maker who is looking for the best-ranked candidate out of a large pool of applicants should plan to interview roughly 50.65\% of them on average.

\section{Full-Information Setting}\label{Sec5}

We now turn to the full-information settings of \cite{Moser1956} and \citet[\S 3]{GilbertMosteller1966}. The optimal solution to these problems is characterized by a decision rule that depends only on the distribution of talent and the number of applicants left to be interviewed. As observed by \cite{GilbertMosteller1966}, no sampling or experience buildup is required in the full-information settings because the aspiration level can be set immediately.

Unfortunately, deriving exact algebraic formulas that describe for any pool size $N$ the expected or median number of interviewed applicants in the full-information settings of \cite{Moser1956} and \citet[\S 3]{GilbertMosteller1966} remains an open problem. In these settings, we can either rely on numerical or asymptotic approximations.

\subsection{Mean and Median Number of Interviews in the Moser (1956) Setting}

Let $A_m=P_i$ refer to the cutoff point when there are $m=N-i$ applicants left to be observed or, stated differently, for the $i\textsuperscript{th}$ applicant observed by the decision-maker in the \cite{Moser1956} setting. As before, let $T$ be the number of applicants interviewed before one is hired.

Keeping in mind that the payoff values observed sequentially by the decision-maker are drawn from a uniform distribution, the probability that no candidate before the $x\textsuperscript{th}$ applicant will have a value larger than its corresponding cutoff point is $\prod_{i=1}^{x-1} P_i$. Since the $x\textsuperscript{th}$ applicant will have a value larger than its corresponding cutoff point $P_x$ with probability $(1-P_x)$, the \emph{ex ante} probability that the $x\textsuperscript{th}$ applicant will be selected by the decision-maker in the \cite{Moser1956} setting is:
\begin{equation}\label{eqProbDist}
    \Pr(T=x) = \left( 1-P_x \right) \prod_{i=1}^{x-1} P_i
\end{equation}

In turn, the expected number of interviews in that full-information setting is:
\begin{equation}\label{eqMeanFullInfo}
\operatorname{\mathbf{E}}(T) = \sum^{N}_{x=1} \left\{ x \Pr(T=x) \right\} = \sum^{N}_{x=1} \left\{ x \left( 1-P_x \right) \prod^{x-1}_{i=1} P_i \right\}
\end{equation}

As long as the cutoff points (also known as the critical values, decision values, minimum acceptable values, or optimal stopping thresholds) can be estimated with a sufficient degree of precision, Eq.~\eqref{eqProbDist} and Eq.~\eqref{eqMeanFullInfo} can be used to evaluate numerically the expected and median number of interviews for any given pool size $N$ in the \cite{Moser1956} setting. This approach has the potential to be especially useful and practical with smaller applicant pools for which the asymptotic limit of $N/3$ derived by \cite{MazalovPeshkov2004}, for example, is not a satisfactory approximation.

Asymptotic results become handy when a decision-maker is dealing with a large pool of applicants. They can also provide useful insights into the type of statistical distribution the decision-maker is dealing with.

\subsubsection{Asymptotic Theory for Moser (1956)}

At this time, it is useful to make use of the asymptotically optimal rule proposed by \cite{BrussFerguson1993} based on the cutoff points:
\begin{equation*} %% \label{eqMoserAmApprox}
    P_i \simeq 1 - \frac{2}{N+2-i} \quad \text{for } i=1,\dots,N
\end{equation*}

The probability distribution of $T$ then telescopes from Eq.~\eqref{eqProbDist} to yield:
\begin{equation*} %% \label{eqProbDistApprox}
    \Pr(T=x) = \frac{2(N+1-x)}{N(N+1)} \quad \text{for } x=1,\dots,N
\end{equation*}

\begin{figure}
\centering
\includegraphics{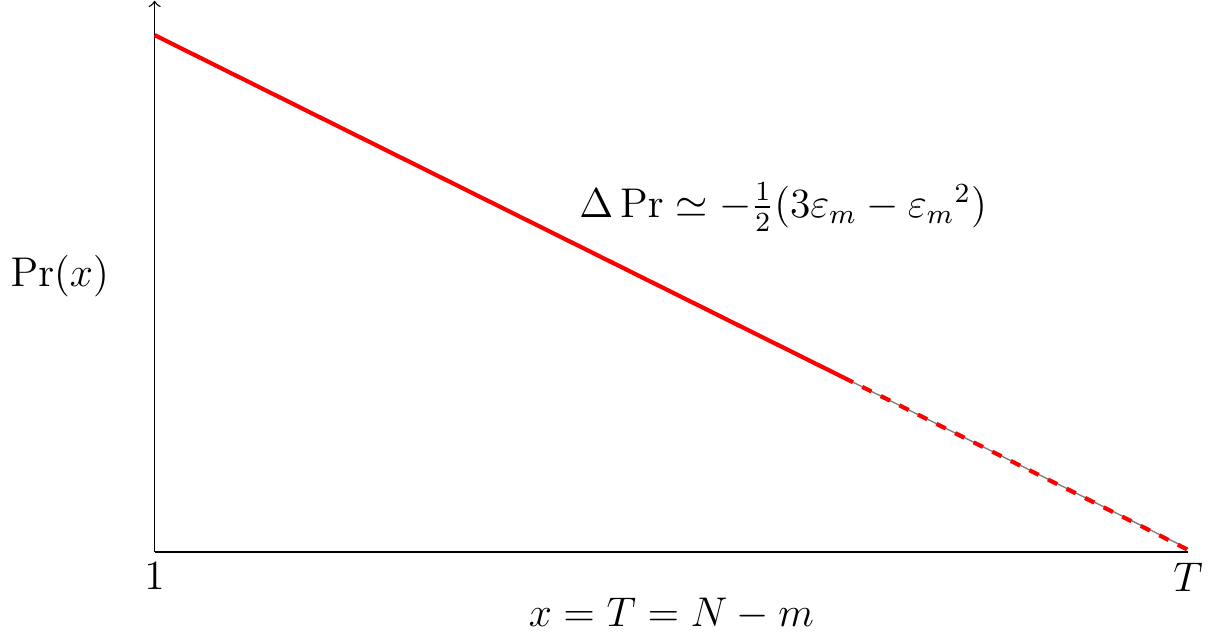}
%% \begin{tikzpicture}[scale=7,font=\large]
%%     \draw[step=0.05,ultra thin,color=gray] (0,0) grid (1.5,0.8);
%%     \draw (0,0) -- (1.5,0);
%%     \draw[->] (0,0) -- (0,0.8);
%%     \draw[very thick,color=red,domain=0:1] plot[id=Prx] function{0.75-0.497*x};
%%     \draw[color=gray,domain=1:1.5] plot[id=Prx2] function{0.75-0.497*x};
%%     \draw[dashed,very thick,color=red] (1,0.252967702) -- (1.05,0.228066745) -- (1.1,0.203159813) -- (1.15,0.178245334) -- (1.20,0.153321018) -- (1.25,0.128383340) -- (1.30,0.103426458) -- (1.35,0.078439575) -- (1.40,0.053399031) -- (1.45,0.028232962) -- (1.5,0.002786836);
%%     \node[right] at (0.6,0.5) {$\Delta \Pr \simeq - \frac{1}{2} ( 3\varepsilon_m - {\varepsilon_m}^2 )$};
%%     \node[left] at (-0.05,0.4) {$\Pr(x)$};
%%     \node[left] at (0,0.75) {$\Pr(x=1) \approx \varepsilon_N$};
%%     \node[below] at (0,0) {$1$};
%%     \node[below] at (1.5,0) {$T$};
%%     \node[below] at (0.75,-0.05) {$x = T = N-m$};
%% \end{tikzpicture}
\caption{Probability distribution of the number of applicants interviewed within the full-information setting of \cite{Moser1956}. The slope is approximately linear for large $m$ (small $x$). Asymptotically, the number of applicants who will be interviewed is asymptotically distributed as a left triangular distribution with lower limit $a=1$, upper limit $b=N$, and mode $c = 1$. The notation $\varepsilon_m$ refers to $2/[m+\ln(m)+1.76799]$.}
\label{Fig2}
\end{figure}

This probability distribution function corresponds approximately to a triangular distribution with lower limit and mode $a=c=1$, and upper limit $b=N$. The argument is illustrated by Fig.~\ref{Fig2}. %% Such a probability distribution yields a mean of $(a+b+c)/3=(N+2)/3$.

It follows that the \emph{proportion} of applicants interviewed by the decision-maker in the \cite{Moser1956} setting is asymptotically distributed as a left triangular distribution with lower limit $a \to 0$, upper limit $b=1$, and mode $c \to 0$. We note in passing that this is a special case of the Beta distribution with parameters $\alpha=1$ and $\beta=2$ \citep[\S 1]{KotzDorp2004}.

It is well known that in such a case the mean is $(a+b+c)/3=1/3$ and the median is $b-\sqrt{(b-a)(b-c)/2} = 1-\sqrt{0.5} \approx 0.292893$ \citep{Ayyangar1941}. In other words, a decision-maker in the \cite{Moser1956} setting who deals with a large pool of applicants should plan to interview one third of those applicants on average, and at least (no more than) 29.29\% roughly half the time.

These theoretical predictions agree with the earlier finding reported by \cite{MazalovPeshkov2004}, as well as numerical estimates. In the presence of $N=10^7$ applicants, for example, the decision-maker in the \cite{Moser1956} setting can expect to conduct approximately 3,333,339 interviews on average. With $N=10^6$ applicants, the decision-maker is equally likely to conduct more or less than 292,897 interviews. With $N=10^7$ applicants, the median is 2,928,936 interviews.

%% Substituting this probability distribution into Eq.~\eqref{eqMeanFullInfo} and dividing by $N$ gives the desired asymptotic result:
%% \begin{equation}\label{eqMoserApprox}
%%     \frac{\operatorname{\mathbf{E}}(T)}{N} = \frac{1}{N} \sum^{N}_{x=1} \left\{ \frac{2x(N-x+1)}{N(N+1)} \right\} = \frac{N+2}{3N} \simeq \frac{1}{3}
%% \end{equation}

%% The formula for the median can be derived similarly. By definition, the asymptotic median in the \cite{Moser1956} will the smallest integer $T$ such that:
%% \begin{equation}
%%     \sum_{x=1}^{T} \frac{2(N+1-x)}{N(N+1)} = \frac{T(2N+1-T)}{N(N+1)} \geq 0.5
%% \end{equation}
%% which yields $\widetilde{T} = (N+\frac{1}{2})(1-\sqrt{0.5})$.

%% This tends to suggest that the median proportion of interviews in the \cite{Moser1956} setting converges asymptotically to:
%% \begin{equation}\label{eqMedianMoser1956}
%% 	\lim_{N \to \infty} \frac{\widetilde{T}}{N} \simeq (1-\sqrt{0.5}) \approx 0.292893
%% \end{equation}

%% These results and the earlier finding reported by \cite{MazalovPeshkov2004} would be consistent with the fact that the proportion of applicants who will be interviewed in the \cite{Moser1956} setting is asymptotically distributed as a left triangular distribution with lower limit $a \to 0$, upper limit $b=1$, and mode $c \to 0$.

Another heuristic argument can be made geometrically. Using the exact recurrence relation described by Eq.~\eqref{eqMoserAm}, we find the slope of the probability distribution in Eq.~\eqref{eqProbDist} as a function of the cutoff points themselves:
\begin{align}\label{eqPrSlope}  %% Slope = m = (Y2-Y1)/(X2-X1) --> If (X2-X1)=1, Y2/Y1 = m+1.
    \Delta \Pr(x) = \frac{\Pr(T=x+1)}{\Pr(T=x)} - 1
    & = \frac{( 1-P_{x+1} ) P_x}{( 1-P_x )} - 1\nonumber\\
    & = \frac{(1-{P_x}^2) P_x}{2(1-P_x)} - 1\nonumber\\
%%    & = - \frac{1}{2} \left( \frac{2(1-P_x) - (1-{P_x}^2) P_x}{(1-P_x)} \right)\nonumber\\
%%    & = - \frac{1}{2} \left( 2 - \frac{(1-{P_x}^2) P_x}{(1-P_x)} \right)\nonumber\\
%%    & = - \frac{1}{2} \left( 2 - \frac{P_x-{P_x}^3}{(1-P_x)} \right)\nonumber\\
%%    & = - \frac{1}{2} \left( 2 - \frac{(1-P_x)({P_x}^2+P_x)}{(1-P_x)} \right)\nonumber\\
    & = - \frac{1}{2} ( 2 - {P_x}^2 - P_x),
\end{align}
%% This uses the fact that:
%% \begin{equation}
%%     1-P_{x+1} = 1-A_{m+1} = 1 - \frac{{A_m}^2+1}{2} = \frac{1-{A_m}^2}{2} = \frac{1-{P_x}^2}{2}
%% \end{equation}

Now, let $\varepsilon_m \approx 2/[m+\ln(m)+1.76799]$ so that $P_x = A_m \simeq 1 - \varepsilon_m$ when $m$ is sufficiently large \citep{GilbertMosteller1966}. (As a reminder, $m$ is the number of applicants left to be observed.) Asymptotically, the slope given by Eq.~\eqref{eqPrSlope} is then:
\begin{equation*}
    \Delta \Pr \simeq - \frac{1}{2} ( 3\varepsilon_m - {\varepsilon_m}^2 )
    %% Wolfram: -3/(x+Log[x]+1.76799) + 2/(x+Log[x]+1.76799)^2
\end{equation*}

This slope is essentially constant for even moderately large values of $m$, in the sense that:
\begin{align*}
    \frac{\partial \Delta \Pr}{\partial x} = - \frac{\partial \Delta \Pr}{\partial m}
    & = - \left( 1+\frac{1}{m} \right) \left( 0.75{\varepsilon_m}^2 - 0.5{\varepsilon_m}^3 \right) \nonumber\\
    & = - 3 \cdot \frac{(m+1)}{m} \cdot \frac{[m+\ln(m)+1.76799-4/3]}{[m+\ln(m)+1.76799]^3} \nonumber\\
    & \simeq - \frac{3}{[m+\ln(m)+1.76799]^2} \nonumber\\
    & \approx 0
\end{align*}

The probability distribution function in Eq.~\eqref{eqPrSlope} therefore might as well be a downward-sloping linear function when there are still many applicants yet to be interviewed (i.e. $m$ is large). When many applicants have been interviewed and the number of applicants left to be observed becomes relatively small (i.e. $m$ is much smaller than $N$), the probability distribution in Eq.~\eqref{eqProbDist} is not quite linear and becomes steeper as $x \to N$. For a large pool size $N$, however, the overall contribution of this extreme right tail will be small (insignificant) relative to the left tail of the probability distribution. Hence, we obtain approximately a triangular distribution.

\subsubsection{Moser (1956) with Decaying Applicant Pool}

One possible variation of the full-information setting of \cite{Moser1956} is to assume that the applicant pool decays or declines over time, in the sense that the single \emph{best} applicant leaves the competition after each successive interview. Mathematically, each independent observation $X_i$ is assumed to be drawn from a (non-identical) uniform distribution over the interval $[0,N+1-i]$ \cite[\S 2, Exercise~9]{Ferguson2008}.

As before, it makes sense to always accept the last observation $X_N$ if we get that far, so $A_0 = 0$. By assumption, that last observation is uniformly distributed over the interval $[0,1]$ and it is therefore expected to yield $\operatorname{\mathbf{E}}(X_{N-1})=1/2$, so we accept the second before last observation $X_{N-1}$ if it is at least $A_1 = 0.5$. More generally, when there are $m=N-i$ applicants left to be observed (``interviewed''), the optimal stopping rule consists in stopping and choosing the $i\textsuperscript{th}$ observation whenever it is larger than the following cutoff point defined recursively (inductively):
\begin{equation*}
    B_{m} = \frac{1}{2} \left( m + \frac{B^2_{m-1}}{m} \right).
\end{equation*}

Moreover, when the pool of applicants left to be observed is large enough, the cutoff points tend to $B_m \simeq m - \sqrt{2m}$ \cite[\S 2, Exercise~9]{Ferguson2008}.

If we get to the $i\textsuperscript{th}$ observation, it will be rejected with probability $A_{m} = B_{m} / (N+1-i)$. As in the classical \cite{Moser1956} setting, these cutoff probabilities can be used to evaluate Eq.~\eqref{eqProbDist} and Eq.~\eqref{eqMeanFullInfo}. This makes it possible to show that the decision-maker who has access initially to $N=10^6$ applicants would be expected to conduct 707.107 interviews on average. With $N=10^7$ applicants, the mean is 2,236.07 interviews.

This leads us to speculate that the mean number of interviews in the \cite{Moser1956} setting with declining applicant quality converges asymptotically to $\operatorname{\mathbf{E}(T)} \simeq \sqrt{N/2}$. Similarly, the median appears to converge to $ \widetilde{T} \simeq \ln{(2)} \sqrt{N/2}$. These computational results would be consistent with the idea that the number of interviews ($T$) in the \cite{Moser1956} setting with a decaying applicant pool is asymptotically distributed as an exponential distribution with the reciprocal of the rate parameter $\lambda^{-1}=\beta=\sqrt{N/2}$. (This would benefit from a formal proof that we omit and leave as an open problem.)

Of note, these results mean that a decision-maker in the full-information setting of \cite{Moser1956} who has to deal with a decaying applicant pool will be expected to conduct an even shorter search than in the \cite{Bearden2006} setting. As in the \cite{Bearden2006} setting, however, the proportion of interviewed applicants will shrink towards zero as the number of applicants grows.

\subsubsection{House-Selling Problem}

Another way to extend the problem posed by \cite{Moser1956} is to impose search costs, while allowing for an unbounded search horizon. This is the framework outlined by \cite{Sakaguchi1961}. In that case, the cutoff point is time-invariant and the mean and median number of expected observations can therefore be expressed as a closed-form solution.

Imagine that something is for sale. Perhaps this could be a house as in the house-selling problem of \cite{MacQueenMiller1960}, or labor as in the job search problem of \cite{Stigler1962}. Let $X_1,X_2,\dots$ denote the independent and identically distributed offers that arrive sequentially, with no possibility of recall. Assume that these values are drawn randomly from a known uniform distribution scaled on the interval $[0,1]$, with no finite limit on the number of offers that can be considered. Suppose that each new observed offer costs a relative amount $c>0$, which could be interpreted interchangeably as search costs, opportunity costs of searching, ongoing maintenance costs, or inventory carrying costs. By construction, the net payoff for accepting the $k\textsuperscript{th}$ offer is $X_k - kc$. When an offer is received, the dilemma is whether it should be accepted or the search for a better offer should continue.

The case with $c>0.5$ is trivial because it never makes sense in that case to observe more than one offer since the cumulative search costs could never be recovered by any subsequent offer.
%% In fact, even the first offer has a negative expected net payoff when $c>0.5$: $\operatorname{\mathbf{E}}(X_1) - c = 0.5-c < 0$. 
%% $2c>1=\max\{X_i\}$
%% That is, $\operatorname{\mathbf{E}}(X_1) - c = 0.5-c < 0$.

When $c \leq 0.5$, the optimal rule is to accept the first offer greater than or equal to the cutoff point $\gamma^* = 1-\sqrt{2c}$ , where $\gamma^*$ is the solution to
\begin{equation*}
    \int_{\gamma}^{1} (x-\gamma) dx = \frac{(1-\gamma)^2}{2} = c.
\end{equation*}

Now, let $T$ represent the number of offers that are considered before the asset (e.g. house) is finally sold. The \emph{ex ante} probability that the first $k-1$ offers will be below the reserve value $\gamma^* = 1-\sqrt{2c}$ and the $k\textsuperscript{th}$ offer will achieve the reserve value is $\Pr{(T=k)} = (1-\sqrt{2c})^{k-1} \sqrt{2c}$. This probability distribution function describes a geometric distribution with success probability $p=\sqrt{2c}$. The expected number of offers to be observed is therefore $\operatorname{\mathbf{E}(T)} = 1/p = 1/\sqrt{2c}$ and the median number of offers would be the solution to $1-(1-\sqrt{2c})^x = 0.5$, or $\widetilde{T} = \lceil \ln{0.5} / \ln{( 1-\sqrt{2c} )} \rceil = \lceil -1 / \log_2{(1-\sqrt{2c})} \rceil$.
%% \begin{equation*}
%% \operatorname{\mathbf{E}}(T) = \sum_{k=1}^{\infty} k \left( 1-\sqrt{2c} \right)^{k-1} \sqrt{2c} = \frac{1}{\sqrt{2c}}.
%% \end{equation*}

%% Similarly, it can be shown that the median number of offers would be the solution to $1-(1-\sqrt{2c})^x = 0.5$, or
%% \begin{equation*}
%%     \widetilde{T} = \left\lceil \frac{\ln{0.5}}{\ln{( 1-\sqrt{2c} )}} \right\rceil = \left\lceil \frac{1}{-\log_2{(1-\sqrt{2c})}} \right\rceil.
%% \end{equation*}
%% Wolfram Alpha: sum x*(1-x)^(k-1) from k=1 to M >> 1-(1-sqrt(2c))^x = 0.5 >> x = -log(2)/log(1-sqrt(2c)).

For illustration purposes, even a small search cost of $c=0.001$ (0.1\% of the maximum value) per offer would result in a relatively short expected search time of $22.36$ offers and a median search time of only $16$ offers. A search cost of $c=0.01$ (1\%) per offer would shrink the search time down to approximately $7.07$ offers on average, with fewer or more than $5$ offers being equally probable.

\subsection{Mean and Median in Full-Information, Best-Choice Setting}

In the full-information best-choice setting of \cite{GilbertMosteller1966}, the decision-maker will select the the $i\textsuperscript{th}$ applicant only if it is a relatively best candidate (with relative rank 1) and it exceeds its decision number or critical value, $P_i$. The optimal policy is described in more details in subsection~\ref{secFIBC}.

The probability that no candidate will be selected among the first $x+1$ observations is $\Pr{(\text{Reach }x+1)} = \sum_{i=1}^{x} {P_i}^x / x$. The \emph{ex ante} probability that the decision-maker will stop and choose the $x\textsuperscript{th}$ applicant follows naturally:
\begin{align}\label{eqProbDistGM1966}
    \Pr{(T=x)} & = \Pr{(\text{Reach }x)} - \Pr{(\text{Reach }x+1)} \nonumber\\
    & = \sum_{i=1}^{x-1} \frac{{P_i}^{x-1}}{x-1} - \sum_{i=1}^{x} \frac{{P_i}^x}{x}
\end{align}

Since the first applicant ($x=1$) is necessarily the relatively best so far, it will be chosen as long as it is better than its corresponding critical value or decision number. This will occur with probability $\Pr{(T=1)} = 1-P_1$.

The last applicant ($x=N$) is also a special case because there is a chance that the decision-maker in the full-information, best-choice setting could interview all $N$ applicants without finding any acceptable candidate that is relatively best. This will occur if the value of the absolute best applicant turns out to be below its corresponding critical value. The associated probability is:
\begin{equation}\label{eqNoChoiceGM1966}
    \Pr{(\text{No Choice})} = \sum_{i=1}^{N} \frac{{P_i}^N}{N}
\end{equation}

The probability density in Eq.~\eqref{eqProbDistGM1966} and Eq.~\eqref{eqNoChoiceGM1966} can be combined to arrive at the cumulative distribution function:
\begin{align}\label{eqCumDistGM1966}
    F(x) =
    \begin{cases}
    1 - \displaystyle \sum_{i=1}^{x} \frac{{P_i}^x}{x} & \quad \text{for } x=1,\dots,N-1\\
%%    & = 1 - \frac{1}{x} \sum_{i=1}^{x} {\left( 1+\frac{c}{N-i} \right)}^{-x}
    1 & \quad \text{at } x=N
    \end{cases}
\end{align}

The expected number of interviews in the full-information, best-choice setting can also be evaluated in a similar manner:
\begin{align}\label{eqMeanGM1966}
 \operatorname{\mathbf{E}}(T) & = \sum^{N}_{x=1} \left\{ x \Pr(T=x) \right\} + N \cdot \Pr{(\text{No Choice})} \nonumber\\
  & = \sum^{N}_{x=1} \left\{  x\sum_{i=1}^{x-1} \frac{{P_i}^{x-1}}{x-1} - x \sum_{i=1}^{x} \frac{{P_i}^x}{x} \right\} + \sum_{i=1}^{N} {P_i}^N
 \end{align}

As in the \cite{Moser1956} setting, these results can be used to evaluate numerically the mean and median number of interviews. In the full-information best-choice setting, the decision-maker who has access to $N=10^4$ applicants is expected to conduct 5,802 interviews on average and more or less than 5,859 interviews half the time. With $N=10^5$ applicants, the mean is 58,017 interviews and the median is 58,592 interviews. Of course, asymptotic approximations would be useful in cases where the pool of applicants is large.

\subsubsection{Asymptotic Theory for Full-Information, Best-Choice Problem}

It is useful at this point to turn to the asymptotic first-order approximation $A_m \approx 1/(1+c/m)$ proposed by \citet[Eq.~3b-2]{GilbertMosteller1966} when there are $m$ applicants left, which translates into $P_{i}=(N-i)/(N-i+c)$ for the $i\textsuperscript{st}$ applicant observed sequentially. As a reminder, in the traditional full-information best-choice setting, the constant $c=0.804352\dots$ continues to be defined by Eq.~\eqref{eqc}.

In this asymptotic context, we can use the fact that ${P_i} \sim \mathrm{e}^{-c}$ for most applicants and therefore ${P_i}^{x-1} = {P_i}^{-1}{P_i}^{x} \sim \mathrm{e}^c \cdot {P_i}^{x}$. More generally, if we let $t$ be defined as the sequential position of the $x\textsuperscript{th}$ applicant scaled along the interval $\{0,1\}$ such that $x=tN$ and use $\alpha$ to represent the fraction of all interviewed applicants such that $i=\alpha N$ and $(1-\alpha)N$ are left to be interviewed, we obtain:
\begin{equation}\label{eqPxGM1966Asymp}
    P_{\alpha}^{t} = \lim_{N \to \infty} {\left( 1 + \frac{c}{(1-\alpha) N} \right)}^{-tN} = \mathrm{e}^{-ct/(1-\alpha)}
\end{equation}
%% \begin{equation}\label{eqAsympGM1966}
%%    \ln{\left( {P_i}^x \right)} & = x\ln{(P_i)} \nonumber\\
%%    & = x \left[ \ln{(N-i+c)} - \ln{(N-i)} \right] \nonumber\\
%%    & \sim -c,
%% \end{equation}

Asymptotically, the decision-maker can expect to interview all $N$ applicants (and fail to find any viable candidate) almost 19.95\% of the time. This follows from the following simplification for Eq.~\eqref{eqNoChoiceGM1966} offered by \cite{Gnedin1996}:
\begin{align}\label{eqNoChoiceGM1966Asymp}
    \lim_{N \to \infty} \Pr{(\text{No Choice})}
    & = \frac{1}{N} \sum_{i=1}^{N} \left( 1+\frac{c}{N-i} \right)^{-N} \nonumber\\
    & \simeq \int_{1}^{\infty} \frac{1}{x^2{\mathrm{e}}^{cx}} dx \nonumber\\
    & = \frac{1}{{\mathrm{e}}^c} - c \cdot \gamma{(0,c)} = 0.199505\dots,
\end{align}
where $\gamma{(0,c)} = \int_{1}^{\infty} {(x{\mathrm{e}}^{cx})}^{-1} dx \approx 0.308164$.

Substituting Eq.~\eqref{eqNoChoiceGM1966Asymp} into Eq.~\eqref{eqMeanGM1966}, passing to the limit, and replacing the sum by an integral allows us to find the proportion of applicants who are expected to be interviewed asymptotically in the full-information best-choice setting. As in the classical Secretary Problem or Dowry Problem, this proportion is asymptotically indistinguishable from the probability that the single best applicant will be successfully selected by the decision-maker:
\begin{align}\label{eqMeanGM1966Asymp}
    \lim_{N \to \infty} \frac{\operatorname{\mathbf{E}}(T)}{N} & = \int_{1}^{\infty} \left\{ \frac{\mathrm{e}^{c}}{x\mathrm{e}^{cx}} - \frac{1}{x\mathrm{e}^{cx}} \right\} dx + \frac{1}{{\mathrm{e}}^c} - c \cdot \gamma{(0,c)} \nonumber\\
    & = \left( \mathrm{e}^{c} - 1 - c \right) \cdot \gamma{(0,c)} + \frac{1}{{\mathrm{e}}^c} \approx 0.580164 = \Pr(\text{Success})
\end{align}

%% Conditional on one candidate being chosen, however, the expected sequential position of that chosen applicant will be closer to the middle: $\operatorname{\mathbf{E}}(T \, | \, \text{Choose})/N = (0.580164-0.199505)/(1-0.199505) \approx 0.4755$.
%% \begin{equation}
%%    \lim_{N \to \infty} \frac{\operatorname{\mathbf{E}}(T) \, | \, \text{Choose One} }{N} = \frac{0.580164-0.199505}{1-0.199505}
%% \end{equation}

By definition, the median $M$ is such that $F(M) = 0.5$, implying that we need to solve $M=2\sum_{1}^{M} {P_i}^M$. If we let $m=\widetilde{T}/N$ represent the median proportion of interviewed applicants, substitute Eq.~\eqref{eqPxGM1966Asymp} asymptotically and replace the sum by an integral, $m$ is defined implicitly as the solution to:
\begin{equation}\label{eqMedianGM1966Asymp}
    m = 2 \int_{0}^{m} \frac{1}{\mathrm{e}^{cm/(1-x)}} dx.
\end{equation}

While deriving a closed-form expression for the median in the full-information best-choice setting remains an open problem, Eq.~\eqref{eqMedianGM1966Asymp} can be solved numerically to yield $m \approx 0.585926$. We note in passing that this theoretical prediction agrees with the numerical estimates presented above.

To summarize, a decision-maker in the full-information best-choice setting of \citet[\S 3]{GilbertMosteller1966} who deals with a large pool of applicants should plan to interview at least (no more than) 58.59\% of those applicants roughly half the time. On average, the decision-maker will interview almost 58.02\% of the applicants. In fact, the decision-maker will interview all applicants without finding any acceptable candidate 19.95\% of the time.
%% In cases where a candidate is chosen, the choice will take place after 47.55\% of all applicants have been interviewed on average.

\subsubsection{Full-Information Duration Problem}

In the full-information version of the duration problem treated by \cite{FergusonHardwickTamaki1992}, it is assumed that each applicant's actual value is revealed to the decision-maker and is sampled from a uniform distribution scaled on the interval $\left[ 0,1 \right]$. As a reminder, the decision-maker seeks to choose a relatively best candidate with the view of maximizing how long it takes before another relatively best (better) candidate is subsequently interviewed, or all the applicants have been interviewed. As in the best-choice problem, it only makes sense for the decision-maker to choose a relatively best candidate.

The solution to this full-information duration problem will have the same form as the full-information best-choice problem: the optimal decision rule for the $i\textsuperscript{th}$ will only depend on the number of applicants left to be observed, and whether it is a relatively best candidate in the first place. It follows that the earlier asymptotic expressions derived in the context of the full-information best-choice problem will continue to apply to the duration problem -- only the problem-specific numerical constant $c$ will be different.

When there are $m$ applicants left and the size of the applicant pool $N$ is large, \cite{FergusonHardwickTamaki1992} showed that the optimal decision value in the full-information duration problem converges to $A_m \simeq 1-2.1198/m$. This is asymptotically equivalent to $P_i \approx (N-i)/(N-i+2.1198)$, where $c=2.1198$ is the implicit solution to:
\begin{equation}
    \mathrm{e}^{c}-1 = \mathrm{e}^{c} \int_{-c}^0 \frac{\mathrm{e}^x}{x} dx - \int_{0}^{c} \frac{\mathrm{e}^x}{x} dx.
\end{equation}

As a side note, precisely the same asymptotic decision rule applies in the full-information best-choice setting of \cite{Porosinski1987}, where the size of the applicant pool is not known precisely but is assumed to be randomly distributed uniformly over the interval $\{1,\dots,b\}$.

In the context of the full-information duration problem with $c=2.1198$, $\gamma{(0,c)} = \int_{1}^{\infty} {(x{\mathrm{e}}^{cx})}^{-1} dx \approx 0.041533$. Substituting into Eq.~\eqref{eqNoChoiceGM1966Asymp} yields the probability that the decision-maker will fail to find any viable candidate asymptotically: $\Pr{(\text{No Choice})} \approx 0.032175$.
%% \begin{equation}
%%    \Pr{(\text{No Choice})} \sim \frac{1}{{\mathrm{e}}^c} - c \cdot \gamma{(0,c)} = 0.032175\dots
%% \end{equation}
As before, Eq.~\eqref{eqMeanGM1966Asymp} gives the expected proportion of interviewed candidates:
\begin{equation}
    \frac{\operatorname{\mathbf{E}}(T)}{N} \simeq \left( \mathrm{e}^{c} - 1 - c \right) \cdot \gamma{(0,c)} + \frac{1}{{\mathrm{e}}^c} = 0.336134\dots
\end{equation}

We note in passing that these last two results agree with those reported by \citet[Theorem~3.1]{MazalovTamaki2006}. %% , with more efforts and a longer proof

Finally, substituting $m = 0.279642$ along with with $c=2.1198$ into Eq.~\eqref{eqMedianGM1966Asymp} is sufficient to show that it is a good asymptotic approximation of the median proportion of applicants who will be interviewed in the context of the full-information duration problem.

\subsubsection{Full-Information Best-Choice Duration Problem}

The corresponding results for the full-information best-choice duration problem can be derived similarly. In the best-choice variant, the optimal decision rule is characterized asymptotically by the constant $c \approx 1.25643\dots$ \citep{FergusonHardwickTamaki1992}, which is the solution to $\mathrm{e}^c=1+2c$. We therefore get $\gamma{(0,c)} \approx 0.144948$, $\Pr{(\text{No Choice})} \simeq 0.10255$, $\operatorname{\mathbf{E}}(T)/N \simeq 0.466785$, and $\widetilde{T}/N \simeq 0.42689$ asymptotically.

\subsection{Numerical Approximations (Finite N)}

Numerical results for the full-information settings of \cite{Moser1956} and \cite{GilbertMosteller1966} are summarized in Table~\ref{Tab2}. The corresponding theoretical predictions in the \cite{Bearden2006} setting and in the context of the Sultan's Dowry Problem \citep{Mosteller1987} are included for comparison purposes.

\begin{landscape}

\begin{table}
\centering
\begin{tabular}{r | r r r r | r r | r r | r r}
 Payoff & \multicolumn{6}{c|}{Cardinal Payoffs} & \multicolumn{4}{c}{Best Choice}\\
 Info. & \multicolumn{4}{c|}{Full Information} & \multicolumn{2}{c|}{Best So Far} & \multicolumn{2}{c|}{Best So Far} & \multicolumn{2}{c}{Full Information}\\
 \hline
 Pool Size & \multicolumn{4}{c|}{\cite{Moser1956}} & \multicolumn{2}{c|}{\cite{Bearden2006}} & \multicolumn{2}{c|}{\cite{Mosteller1987}} & \multicolumn{2}{c}{\citetalias[\S 3]{GilbertMosteller1966}}\\
 ($N$) & $\operatorname{\mathbf{E}}(T)$ & $\operatorname{\mathbf{E}}(T) / N$ & $\widetilde{T}$ & $\widetilde{T} / N$ & $\operatorname{\mathbf{E}}(T)$ & $\widetilde{T}$ & $\operatorname{\mathbf{E}}(T)$ & $\widetilde{T}$ & $\operatorname{\mathbf{E}}(T)$ & $\widetilde{T}$\\
 \hline  \hline
 9 & 4.23844 & 0.4709 & 3 & 0.3333 & 5.68571 & 4 & 7.02857 & 6 & 5.50788 & 5\\
 25 & 9.87275 & 0.3949 & 8 & 0.3200 & 11.9372 & 8 & 18.8979 & 18 & 14.7839 & 14\\
 49 & 18.0805 & 0.3690 & 15 & 0.3061 & 19.1778 & 12 & 36.7214 & 36 & 28.7061 & 28\\
 64 & 23.1646 & 0.3619 & 20 & 0.3125 & 23.0590 & 14 & 47.2265 & 46 & 37.4082 & 37\\
 100 & 35.3069 & 0.3531 & 30 & 0.3000 & 31.2266 & 18 & 74.4780 & 74 & 58.2935 & 58\\
 400 & 135.758 & 0.3394 & 118 & 0.2950 & 77.4217 & 38 & 294.837 & 294 & 232.342 & 234\\
 2500 & 836.365 & 0.3345 & 734 & 0.2936 & 242.191 & 98 & 1840.38 & 1840 & 1513.68 & 1452\\
 10000 & 3336.83 & 0.3337 & 2931 & 0.2931 & 556.413 & 198 & 7358.48 & 7358 & 5802.01 & 5859\\
 1000000 & 333338 & 0.3333 & 292897 & 0.2929 & 7901.35 & 1998 & 735759 & 735758 & 580165 & 585925\\
%% Mosteller (1987) Threshold = \approx Wolfram Alpha: N/e^(1+1/(2*N)) OR ceiling(-1/(2*LambertW(-(e^(1-1/(2*N))/(2*N)))))
%% UPDATED: Numerical estimates for GM (1966) setting. Now relying on (almost) 256 bits of precision. Exact up to N=2500. Asymptotic estimate for P_i used to calculate N=10000 and N=1000000.
 \hline
\end{tabular}
\caption{Expected and median number of interviews for finite pool of applicants. The decision-maker observes the uniformly distributed payoff values \citep{Moser1956,GilbertMosteller1966} or only observes whether an applicant is the best so far \citep{Bearden2006}, or only cares about picking the best applicant anyway \citep{Mosteller1987}.}
\label{Tab2}
\end{table}

\end{landscape}

Table~\ref{Tab2} reveals a few stylized facts. First, for a given pool size ($N$) and informational setting, the decision-maker who only cares about picking the best applicant is expected to conduct more interviews compared to the decision-maker who is rewarded based on the actual intrinsic value of the selected candidate. Moreover, in a best-choice setting, the decision-maker who only receives a rank-based signal \citep{Mosteller1987} is expected to conduct more interviews for a given pool size compared to the decision-maker who can observe the actual payoff values \citep{GilbertMosteller1966}. We note in passing that this is not strictly true in a setting with cardinal payoffs. As long as there are fewer than $N = 64$ available applicants under consideration, \emph{fewer} interviews are expected to take place when the payoff values are fully observable by the decision-maker \citep{Moser1956} compared to when the decision-maker only receives a rank-based signal of relative attractiveness \citep{Bearden2006}. However, as soon as the pool of applicants grows beyond 64 or so choices, the decision-maker who can observe the actual payoff values will be expected to conduct \emph{more} interviews on average in a setting with cardinal payoffs.

\subsection{Summary of Asymptotic Results (Large N)}

For greater clarity and to put our earlier findings in context, Table~\ref{Tab3} summarizes asymptotic results that can be used to estimate the mean and median number of interviews in the full-information settings of \cite{Moser1956} and \citet[\S 3]{GilbertMosteller1966}, where the decision-maker can observe the uniformly distributed payoff values, as well as the no-information settings of \cite{Mosteller1987} and \cite{Sakaguchi1984}, where the decision-maker only receives a signal of relative rank. Specifically in the \cite{Sakaguchi1984} setting, the decision-maker also has the option to not pick any applicant even after interviewing all of them. The rank problems of \cite{Lindley1961} and \cite{GuseinZade1966} are also represented, along with the duration problem of \cite{FergusonHardwickTamaki1992}.

%% As a reminder, the decision-maker only cares about picking the best possible applicant in the \cite{Mosteller1987}, \cite{GilbertMosteller1966} and \cite{Sakaguchi1984} settings. In the \cite{PresmanSonin1972} setting, the exact number of available applicants ($N$) is not known precisely but is assumed to be uniformly distributed over the known interval $\{1,\dots,b\}$.

\begin{landscape}

\begin{table}
\centering
\begin{tabular}{l l l r r}
 \hline
 Setting & Info. & Payoffs & Mean ($\bar{T}/N$) & Median ($\widetilde{T}/N$) \\  %% Min. & 
 \hline  \hline
 \cite{FrankSamuels1980} & RR & Best $S \to \infty$
  & $t^* \approx 0.2834$
  & $t^* \approx 0.2834$ \\
 \cite{Moser1956} & FI & Cardinal
  & $1/3 \approx 0.3333$
  & $1-\sqrt{0.5} \approx 0.2929$ \\  %% Section~\ref{Sec5}
 \cite{FergusonHardwickTamaki1992} & FI & Duration of BSF
  & $\pi(2.1198) \approx 0.3361$
  & $m(2.1198) \approx 0.2796$ \\
 & BSF & Duration of BSF  %% \cite{FergusonHardwickTamaki1992}
  & $3/\mathrm{e}^2 \approx 0.4060$
  & $2/\mathrm{e}^2 \approx 0.2707$ \\
 & FI & Duration $\times$ BC  %% \cite{FergusonHardwickTamaki1992}
  & $\pi(1.25643) \approx 0.4668$
  & $m(1.25643) \approx 0.4269$ \\
 \cite{GuseinZade1966} & RR & Best $S=25$
  & 0.4700
  & 0.4555 \\
 \cite{Lindley1961} & RR & Rank
  & 0.5065
  & $\sqrt{2} \sqrt[4]{3} / V_{\infty} \approx 0.4810$ \\
 \cite{GuseinZade1966} & RR & Best $S=15$
  & 0.5095
  & 0.4904 \\
 \cite{FergusonHardwickTamaki1992} & BSF & Duration $\times$ BC
  & $- \frac{3}{2}\mathbf{W}(*) - \frac{1}{2}\mathbf{W}(*)^{2} \approx 0.5270$
  & $- \mathbf{W}(*) \approx 0.4064$ \\
 \citetalias[\S 3]{GilbertMosteller1966} & FI & BC $\{0,1\}$
  & $\pi(0.804352) \approx 0.5802$
  & $m(0.804352) \approx 0.5859$ \\
 \cite{GuseinZade1966} & RR & Best $S=5$
  & 0.6102
  & 0.5771 \\
%% \citetalias{PresmanSonin1972} & BSF, $N \sim \mathrm{U}(1,b)$ & BC $\{0,1\}$
%%  & $\left[4-2\ln(2)\right]/\mathrm{e}^{4} \approx 0.6244$
%%  &  \\  %% $4/\mathrm{e}^{2} \approx 0.5413$
 & RR & Best $S=3$  %% \cite{GuseinZade1966}
  & 0.6564
  & 0.6286 \\
 & RR & Best $S=2$  %% \cite{GuseinZade1966}
  & 0.6892
  & 0.6802 \\
 \cite{Mosteller1987} & BSF & BC $\{0,1\}$
  & $2/\mathrm{e} \approx 0.7358$
  & $2/\mathrm{e} \approx 0.7358$ \\
 \cite{Szajowski1982} & RR & $2\textsuperscript{nd}$ Best $\{0,1\}$
  & $1/2+\ln(2)/2 \approx 0.8466$
  & 1 \\
 \cite{Sakaguchi1984} & BSF & BC $\{-1,0,1\}$
  & $(3/2)/\sqrt{\mathrm{e}} \approx 0.9098$
  & 1 \\
 \hline
\end{tabular}\\
FI: Full Information. BSF: Best So Far Indicator. RR: Relative Rank. BC: Best Choice.\\
The notation $\mathbf{W}(*)$ stands for the Lambert W function with argument $c^*/N = -2 / \mathrm{e}^{2}$.\\
$\pi(c) = \left( \mathrm{e}^{c} - 1 - c \right) \gamma{(0,c)} + {\mathrm{e}}^{-c}$, with $\gamma{(0,c)} = \int_{1}^{\infty} {(x{\mathrm{e}}^{cx})}^{-1} dx$.\\
$m(c)$ is taken to be the solution to $m = 2 \int_{0}^{m} \exp{(-\frac{cm}{1-x})}$.\\[5pt]
\caption{Asymptotic mean and median proportion of interviewed applicants ($N \to \infty$).}
\label{Tab3}
\end{table}

\end{landscape}

Perhaps coincidentally, the median proportion of interviewed applicants expected in the full-information best-choice setting of \cite{FergusonHardwickTamaki1992} is close to $2-2/\sqrt{2} \approx 0.5858$, which is precisely twice the median proportion of interviewed applicants in the full-information setting with cardinal payoffs of \cite{Moser1956}. Similarly, the median proportion of interviewed applicants expected in the \cite{Moser1956} setting is \emph{approximately} equal to the median proportion of interviewed applicants in the \cite{Mosteller1987} setting to the fourth power. That is, $(2/\mathrm{e})^4 \approx 0.293 \approx 1-\sqrt{0.5}$.

\section{Conclusion}

Optimal stopping problems give rise to random distributions describing how many interviews might be conducted by the decision-maker. Despite the fact that they have practical implications, these probability distributions are rarely studied. This review focuses on the problem of choosing a candidate from a pool of applicants with uniformly distributed talent. Asymptotically, the proportion of applicants who can expected to be interviewed goes to $x-x\ln{(x)}$, where $x$ is the proportion of applicants who should be passed over according to the optimal threshold rule (Section~\ref{Sec3}). To estimate the mean and median proportion of applicants who can expected to be interviewed in the asymptotic version of the minimum expected rank problem of \cite{Lindley1961}, for example, we can leverage the earlier findings of \cite{Yeo1997}. This is covered in Section~\ref{Sec4}. In terms of theoretical contributions, we show that the number of interviews in the full-information setting of \cite{Moser1956} is asymptotically distributed as a left triangular distribution (Section~\ref{Sec5}). This puts into their proper context the earlier findings of \cite{MazalovPeshkov2004}.

In the tongue-in-cheek conclusion of his rejoinder, \cite{FergusonRejoinder1989} equates marriage as the outcome of an optimal stopping problem and wonders out loud when it would become optimal for people to get married. In that context, the duration of the search could be interpreted as the age when the decision-maker gets into a committed relationship. Table~\ref{Tab3} suggests that only the most lenient \citep{FrankSamuels1980} or most discerning \citep{Moser1956} decision-makers are expected to choose a partner before half of all available options have been sampled. Even when time is of the essence \citep{FergusonHardwickTamaki1992}, marriage would be expected only after more than one third of all options have been sampled on average. Incredibly, marriage on average would be expected to typically take place between people who are in the second half of their life! This is much later than what is hinted by the famous optimal threshold of $1/\mathrm{e} \approx 0.368$. Even in the classical no-information best-choice setting, people would be expected to get married only in the last third of their life on average. In fact, a majority of people in different contexts are not even expected to \emph{ever} get married \citep[for example]{Szajowski1982,Sakaguchi1984}.

\end{document}